\documentclass{jfm_arxiv}

\usepackage{graphicx}
\usepackage{newtxtext}
\usepackage{newtxmath}
\usepackage{natbib}
\usepackage{hyperref}
\hypersetup{
colorlinks = true,
urlcolor = blue,
citecolor = black,
}

\newcommand{\RomanNumeralCaps}[1]
\linenumbers

\usepackage{booktabs}
\usepackage[justification=justified, singlelinecheck=false]{caption}
\usepackage{siunitx}
\usepackage{xcolor}

\title{Assessment of non-intrusive sensing in wall-bounded turbulence through explainable deep learning }

\author{A. Cremades\aff{1},
R. Freibergs\aff{1},
S. Hoyas\aff{2},
A. Ianiro\aff{3},
S. Discetti\aff{3},
\and R. Vinuesa\aff{1} \corresp{\email{rvinuesa@mech.kth.se}}}

\affiliation{\aff{1}FLOW, Engineering Mechanics, KTH Royal Institute of Technology, SE-100 44 Stockholm, Sweden \aff{2} Instituto Universitario de Matemática Pura y Aplicada, Universitat Politècnica de València, Valencia, 46022, Spain \aff{3} Department of Aerospace Engineering, Universidad Carlos III de Madrid, Avenida de la Universidad 30, 28911 Leganés, Madrid, Spain}

\begin{document}
\maketitle

\begin{abstract}
In this work we present a framework to explain the prediction of the velocity fluctuation at a certain wall-normal distance from wall measurements with a deep-learning model. For this purpose, we apply the deep-SHAP method to explain the velocity fluctuation prediction in wall-parallel planes in a turbulent open channel at a friction Reynolds number ${\rm{Re}}_\tau=180$. The explainable-deep-learning methodology comprises two stages. The first stage consists of training the estimator. In this case, the velocity fluctuation at a wall-normal distance of 15 wall units is predicted from the wall-shear stress and wall-pressure. In the second stage, the deep-SHAP algorithm is applied to estimate the impact each single grid point has on the output. This analysis calculates an importance field, and then, correlates the high-importance regions calculated through the deep-SHAP algorithm with the wall-pressure and wall-shear stress distributions. The grid points are then clustered to define structures according to their importance. We find that the high-importance clusters exhibit large pressure and shear-stress fluctuations, although generally not corresponding to the highest intensities in the input datasets. Their typical values averaged among these clusters are equal to one to two times their standard deviation and are associated with streak-like regions. These high-importance clusters present a size between 20 and 120 wall units, corresponding to approximately 100 and 600\si{\micro\meter} for the case of a commercial aircraft.
\end{abstract}


%{\bf MSC Codes } {\it(Optional)} Please enter your MSC Codes here

\section{Introduction}\label{sec:introduction}

The International Energy Agency estimated that around 30\% of the energy consumption worldwide is used for transportation~\citep{IEA2020}, being 15\% of this energy spent near the boundaries of the vehicle~\citep{Jimenez2013}. Therefore, a deeper understanding of fluid mechanics and turbulence will have strong implications in fuel consumption, cost savings, and reducing carbon dioxide emissions worldwide~\citep{kim2011}. Knowing the mechanisms that turbulence uses for transporting energy and generating friction drag is a crucial point for creating new control strategies to face the environmental challenges worldwide~\citep{marusic2021}. However, turbulent flows are described by the Navier--Stokes equations~\citep{navier1827,Stokes2009}, a set of partial differential equations whose analytical solution has not been found for a general case of application. Additionally, energy is transported through different scales, from the largest to the smallest ones, until it is dissipated~\citep{kol41a}, and in the opposite direction~\citep{Cardesa_science}, requiring costly and large simulations~\citep{hoy22,osawa2024} or experimental tests~\citep{tay38,kli67,tow76,deshpande2023}.

The chaotic behavior of turbulent flows does not allow extrapolating the instantaneous deterministic predictions of a certain simulation or experimental test to a real-world application. One possibility to overcome this limitation is the use of non-intrusive sensing to reconstruct the flow from external sensors. Non-intrusive techniques have been widely used for experimental tests in fluid dynamics as they do not require placing any sensor inside the rig. The most common techniques involve the use of laser and optical techniques, such as Laser-Doppler velocimetry~\citep{george1973} or particle image velocimetry~\citep{adrian2011}. However, for real-world applications, these techniques are not suitable as the sensing equipment cannot be installed in the device itself. Thus, new methodologies based, for example, on deep learning~\citep{lecun2015} are required to reconstruct the flows~\citep{cuellar2024}.

In the last decades, machine learning (ML) has been widely used as a result of the large amounts of available data~\citep{chui2018} and the improvement in computational power~\citep{al2015}. Fluid dynamics is not exempt from this trend, as ML has provided novel possibilities for the analysis of flows. ML models have been used to improve the results of numerical simulations~\citep{brunton} and experimental results~\citep{vinuesa2023}: modeling turbulence~\citep{wang2017,beck2019}, accelerating the simulations~\citep{bar2019,li2020,kochkov2021,marino2024}, creating reduced order models~\citep{lee2020,kaiser2014}, improving experimental techniques~\citep{rabault2017}, controlling~\citep{kim2024} and reconstructing flows~\citep{yousif2023}. Related to the flow reconstruction, ML has also been applied to non-intrusive sensing~\citep{lee1997,inigo2014,kim2020}. In these works, wall measurements are used to predict the flow evolution. For instance, \citet{Guastoni2020} estimated the velocity at a certain wall-normal distance inside a turbulent open channel through a deep-learning model. However, in all these works ML is used as a black-box model. The accuracy of the predictions and the statistics of the reproduced flow are analyzed, but the relationships between the inputs and the outputs are not explored. Therefore, a deeper analysis is required to understand the most influential regions for the prediction, if those regions form clusters, and the physical properties of those structures.

Explaining deep-learning models is crucial for understanding the physics of the predicted flow. Additive-feature-attribution methods propose a simplified linear model to understand the relationships between inputs and outputs for a general nonlinear model, capturing the nonlinear interactions between the input features~\citep{cremades2025}. These methods are based on the Shapley values~\citep{shapley1953} and adapted to the architecture of the deep-learning models~\citep{lundberg2017}. Some researchers such as \citet{he2022}, \citet{mcconkey2022} and \citet{sudharsun2023} have applied the additive-feature-attribution methods to understand the effect of the flow parameters in the prediction of the eddy viscosity used in Reynolds averaged Navier--Stokes turbulence modeling~\citep{boussinesq1903}. Others, such as \citet{Lellep2022} or \citet{cremades2024} have used them to explain the evolution of turbulent flows. Applying this methodology to the non-intrusive sensing problem provides new insight into the patterns predicted by the deep-learning model. A proper understanding of them is essential for real-world applications in which the amount of sensors is limited and they need to be located optimally.

This work extends the non-intrusive sensing proposed by \citet{Guastoni2020}. The importance of the measured wall pressure and shear stress for the prediction of the velocity fluctuations at a certain wall distance is evaluated. The impact of each input grid point on the model is calculated by applying an additive-feature-attribution method known as deep SHAP~\citep{lundberg2017}. Then, the grid points are ranked according to their importance for the reconstruction of the flow. Finally, the domain is segmented and the physical properties of the generated coalitions are analyzed.

The manuscript is structured as follows. First, the methodology of the feature attribution methods used for the analysis is presented. The workflow is illustrated, and the underlying mathematical framework for the calculations is thoroughly detailed. Then, the results of the explainable methods are discussed. The statistical distribution of each grid point's importance and its respective wall stress is presented. The connection between the input values and their score in the prediction is unveiled and the sensitivity of the model to each input grid point is evaluated. The domain is segmented into similar-importance coalitions whose properties are calculated and discussed. Finally, the results are summarized, and the main conclusions of the work are presented.

\section{Methods}\label{sec:methods}

The present work evaluates the importance of the wall pressure and wall-shear stress in a turbulent open-channel at a friction Reynolds number $\text{Re}_\tau = \left(u_\tau h\right)/\nu = 180$ to predict the velocity fluctuations at a certain wall-normal distance, $y^+=\left(u_\tau y\right)/\nu=15$. The friction velocity, $u_\tau=\sqrt{\tau_w/\rho}$, is defined based on the shear stress on the wall $\tau_w$ and the fluid density $\rho$ while $\nu$ is the fluid kinematic viscosity. Note that the wall-unit magnitudes are represented by the superscript $^+$, the averaged value of a variable $x$ by $\overline{x}$, and its standard deviation by $\sigma_x$. Finally, the standardized values are calculated as $\widetilde{x}=\left(x-\overline{x}\right)/\sigma_x$.

\begin{figure}
\centering
\includegraphics[width=1\linewidth]{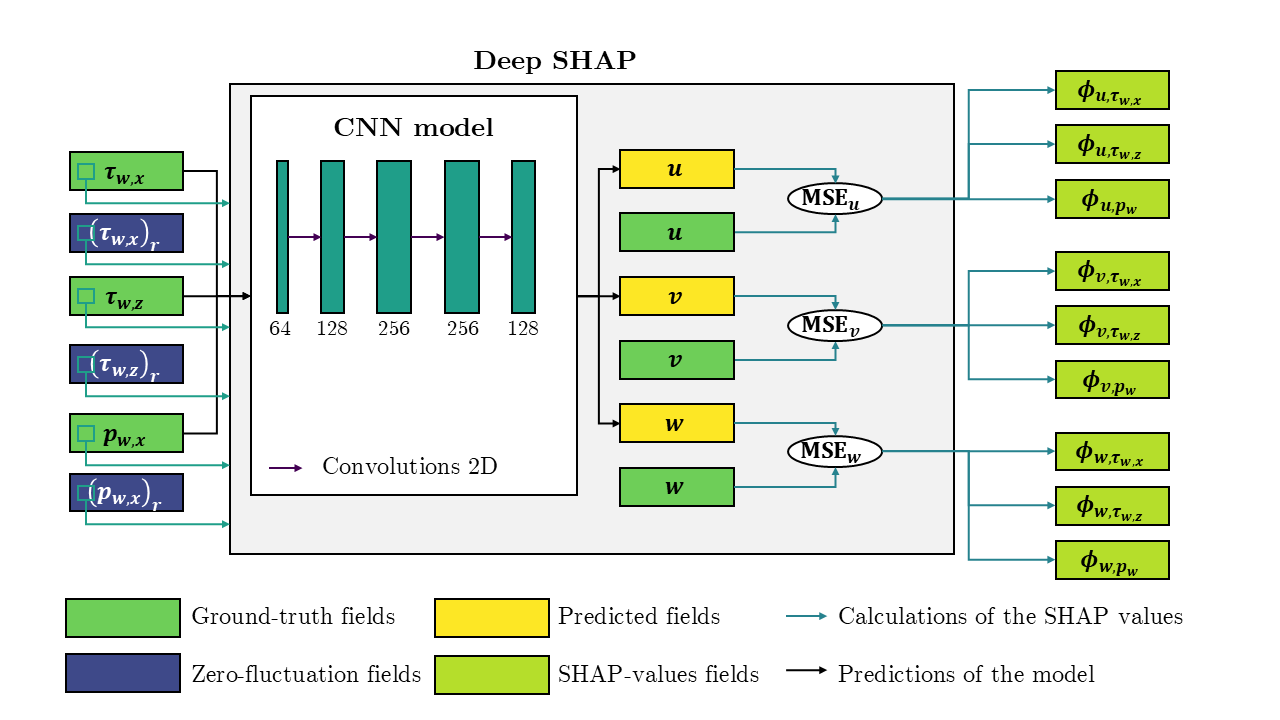}
\caption{Workflow of the methodology. The figure shows the model used for calculating the SHAP values. This model comprises two stages. The first stage corresponds to the prediction of the velocity fluctuation (yellow boxes) at a wall distance $y^+=15$ from the wall measurements (green boxes). The workflow of the velocity prediction is presented with the black arrows. Then, the SHAP values (light green boxes) are calculated using a deep SHAP explainer. First, the non-informative fields are selected (blue boxes), and then, the output of the model is compared to the ground truth (green boxes) through the mean squared error (MSE).}
\label{fig:workflow}
\end{figure}

This work modifies the original deep-learning model of \citet{Guastoni2020} to calculate the importance of the input grid points in the output prediction. The original model predicts the velocity fluctuations from the wall pressure and shear stress using a convolutional neural network ~\citep[CNN,][]{lecun1989}, exploiting its ability to extract information and local correlation in images or volumes~\citep{murata2020,morimoto2021}. Then, to quantify the accuracy of the CNN, an extra layer is included in the model to calculate the mean squared error between the prediction of the original model and the ground truth of each velocity fluctuation. The SHAP (Shapley additive explanations) values are calculated following two stages, prediction of the velocity and evaluation of the errors, which are summarized in Figure \ref{fig:workflow}.

The convolutional neural network is trained using the wall pressure $p_w$, and the wall-shear stress in the streamwise and the spanwise directions $\tau_{wx}$ and $\tau_{wz}$. The model is then used to predict the velocity fluctuations in the streamwise, $u$, wall-normal, $v$, and spanwise, $w$ directions. The input and output fields have $192\times 192$ grid points. A $15$ grid-point periodic padding is added to the input fields to avoid boundary inaccuracies. As reported by \citet{Guastoni2020}, the network is trained to obtain a reconstruction error lower than 1\%.

Then, the deep-SHAP algorithm~\citep{lundberg2017} is used for evaluating the importance of each wall grid point for a total of 1000 fields. Due to the complexity of the deep-learning model, sketched in Figure \ref{fig:workflow}, the relationships between the inputs and the outputs cannot be understood by analyzing the model itself. According to \citet{lundberg2017}, the most interpretable models are those composed of linear relationships. For this reason, a simplified or explainable equivalent model, $g$, needs to be calculated. The additive-feature-attribution methods are based on this explainable model, Equation (\ref{eq:afam}), to exploit the explainability potential of the linear models.

\begin{equation}
\label{eq:afam}
g\left(q\right) = \phi_0+\sum_{i=1}^{N}\phi_i q'_i
\end{equation}

\noindent The additive-feature-attribution methods reduce the original model $f$ to the approximate model $g$ presented in Equation (\ref{eq:afam}). This approximate function is a linear summation of the SHAP values, $\phi_i$, for the input feature $q'_i$, which is a boolean value equal to $1$ if the feature is present and $0$ if it is absent. The value $\phi_0$ equals the function $g$ when all the features are absent, in other words, when the non-informative input is used. These additive feature attribution methods must satisfy three axioms: local accuracy, missingness, and consistency~\citep{lundberg2017}. The only possible solution that satisfies all of them at the same time is the Shapley values~\citep{shapley1953}, equation (\ref{eq:shapley_values}).

\begin{equation}
\label{eq:shapley_values}
\phi_{i} = \sum_{S\subseteq F\backslash\left\{i\right\}}\frac{\left|S\right|!\left(N-\left|S\right|-1\right)!}{N!}\left[f\left(S\cup i\right)-f\left(S\right)\right]\text{.}
\end{equation}

\noindent Here the Shapley values $\phi_i$ of the feature $i$ are calculated as a weighted average of the error in the output model of a certain subset $S$, including and lacking the feature $i$: $\left[f\left(S\cup i\right)-f\left(S\right)\right]$. The subset is taken from the total set $F$ that does not include the feature $i$. The error is averaged by weighting its value with the probability of the subset $S$ to be taken before the feature $i$: $\left(\left|S\right|!\left(N-\left|S\right|-1\right)!\right)/\left(N!\right)$. The parameter $\left|S\right|$ represents the number of not null features in the subset $S$ and $N$ is the total number of subsets. The computation of the Shapley value scales exponentially with the number of input features, $2^N$, making its calculation extremely expensive for deep learning models~\citep{jia2019}.

To reduce the computational requirements of the Shapley values, they can be approximated by applying the knowledge of the model architecture. This work uses the deep-SHAP methodology, a combination of the DeepLIFT~\citep[Learning Important FeaTures,][]{Ribeiro2016} and the Shapley values, for computing the importance of the input features in the output predictions. The SHAP values are calculated assuming independent features and a linear model. DeepLIFT calculates the variation in the output $\Delta o$ as the summation of the attributions, $\sum_{i=1}^{n} C_{\Delta x_{i},\Delta o}$, of each input feature concerning its reference value, $\Delta x_i$: $\sum_{i=1}^{n} C_{\Delta x_{i},\Delta o} = \Delta o$, being $n$ the number of inputs, $\Delta o = f\left(x\right)-f\left(x_{r}\right)$, and $\Delta x_{i} = x-x_{r}$. The attributions of the DeepLIFT method, $C_{\Delta x_{i},\Delta o}$, are equivalent to the SHAP values $\phi_{i}$ and $f\left(x_{r}\right)$ equals $\phi_0$. In the deep SHAP methodology, the SHAP values of the smaller components are calculated approximating linearly the model:

\begin{equation}
\label{eq:linearshap}
\phi_{i}\left(f_{j},x\right) \approx m_{x_{i},f_{j}}\left(x_{i}-\mathbb{E}\left[x_{i}\right]\right)
\end{equation}

\noindent where $f_j$ refers to the component $j$ of the model $f$ and $m_{x_{i},f_{j}}$ is the coefficient or attribution of the feature $x_i$ in the calculation of the output of the component $j$ of the model $f$ and $\mathbb{E}\left[x_{i}\right]$ the expected value of the feature $x_i$. Then, these local SHAP values are recursively passed backward in the network using the chain rule.

\section{Results}\label{sec:results}

\begin{figure}
\includegraphics[width=\textwidth]{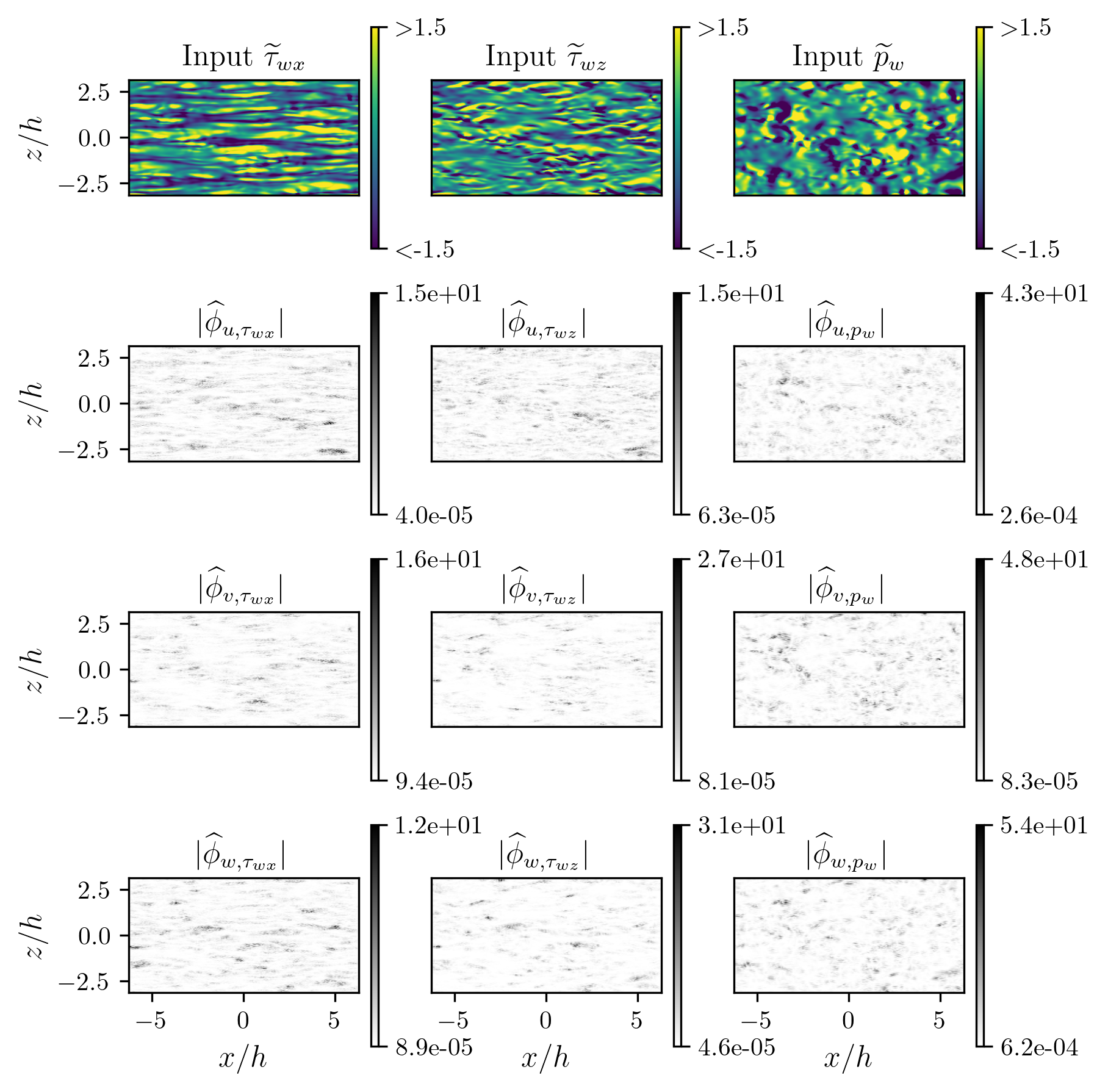}
\caption{Instantaneous visualization of the wall measurements of the open channel (streamwise wall-shear stress $\tau_{wx}$, spanwise wall-shear stress $\tau_{wz}$ and wall-normal pressure $p_w$) and their corresponding SHAP values for the predictions of the velocity fluctuations ($u$, $v$ and $w$).}
\label{fig:Mean1000}
\end{figure}

An importance field, {\it i.e.,} the SHAP values for all the input grid points, is calculated by correlating each input wall stress (streamwise shear stress, spanwise shear stress, and pressure) with its importance in the calculation of the velocity fluctuation in one direction (streamwise, wall-normal, spanwise) at a wall-normal distance $y^+=15$. The SHAP values are calculated by comparing the input fields with a non-informative reference, the mean value of the wall-shear stress and pressure $i$. The larger the magnitude of the SHAP values used for this purpose, the more sensitive the model output is to any change in the input feature.
Figure~\ref{fig:Mean1000} shows the instantaneous normalized SHAP values:

\begin{equation}
\hat{\phi}_{u_i,p_j}=\frac{3 N_x N_z\left\vert \phi_{u_i,p_j}\right\vert}{\sum_{i_x}^{N_x}\sum_{i_z}^{N_z}\left(\phi_{u_i,\tau_{wx}}+\phi_{u_i,\tau_{wz}}+\phi_{u_i,p_w}\right)}\text{,}
\end{equation}

\noindent connecting the pairs of input, $p_j$, and output, $u_i$, variables. Therefore, these images visualize whether and how the importance scores correlate with the input.

%The calculation returns nine arrays of SHAP values with the same size as the inputs, $208\times208$, corresponding to the importance of each input stress in the prediction of an output velocity. For example, how relevant each pixel in the $\tau_{wx}$ input channel is for the prediction of the streamwise fluctuation $u$.

%Note that the plots include the maximum and minimum values of the mean SHAP scores, so the scale of the impact of each input feature can be easily visualized. Note that the mean of the SHAP values and input variables is taken over multiple snapshots, being essentially equivalent to taking an average in time over 1000 samples.

Figure~\ref{fig:Mean1000} shows the instantaneous standarized wall pressure and shear stress:
\begin{equation}
\widetilde{p}_{w}=\frac{p_{w}-\overline{p_{w}}}{\sigma_{p_{w}}}\text{, }
\widetilde{\tau}_{wx}=\frac{\tau_{wx}-\overline{\tau_{wx}}}{\sigma_{\tau_{wx}}}\text{ and }
\widetilde{\tau}_{wz}=\frac{\tau_{wz}-\overline{\tau_{wz}}}{\sigma_{\tau_{wz}}}\text{,}
\end{equation}

\noindent as well as the corresponding SHAP values for predicting the velocity fluctuations at $y^+=15$. The SHAP values detect the high-importance regions of the model, which represents a trajectory of the turbulent flow. Therefore, we can infer that the SHAP values detect the regions of the flow that are most influential for the reconstruction of the fluctuations of the velocity. High-interest regions are those with a high absolute SHAP value. The figure demonstrates that the important regions for the predictions are localized in small clusters of the flow, while most of the grid points exhibit low importance. The visual resemblance between the inputs and their SHAP values might be inferred from the figure. In fact, SHAP values related to the streamwise shear stress at the wall tend to create elongated structures, which are observed more clearly in the prediction of the streamwise velocity due to the strong correlation between both variables~\cite{cuellar2024_2}. Similarities between the input and the SHAP values are also observed for the spanwise shear stress and pressure, being the high-importance clusters less elongated than the corresponding streamwise shear stress SHAP clusters.
In addition, the figure demonstrates that the model is highly sensitive to the pressure variations as the maximum normalized SHAP value for the pressure is higher than for the shear stress in the prediction of a specific component of the velocity fluctuation.

\begin{figure}
\includegraphics[width=0.49\textwidth]{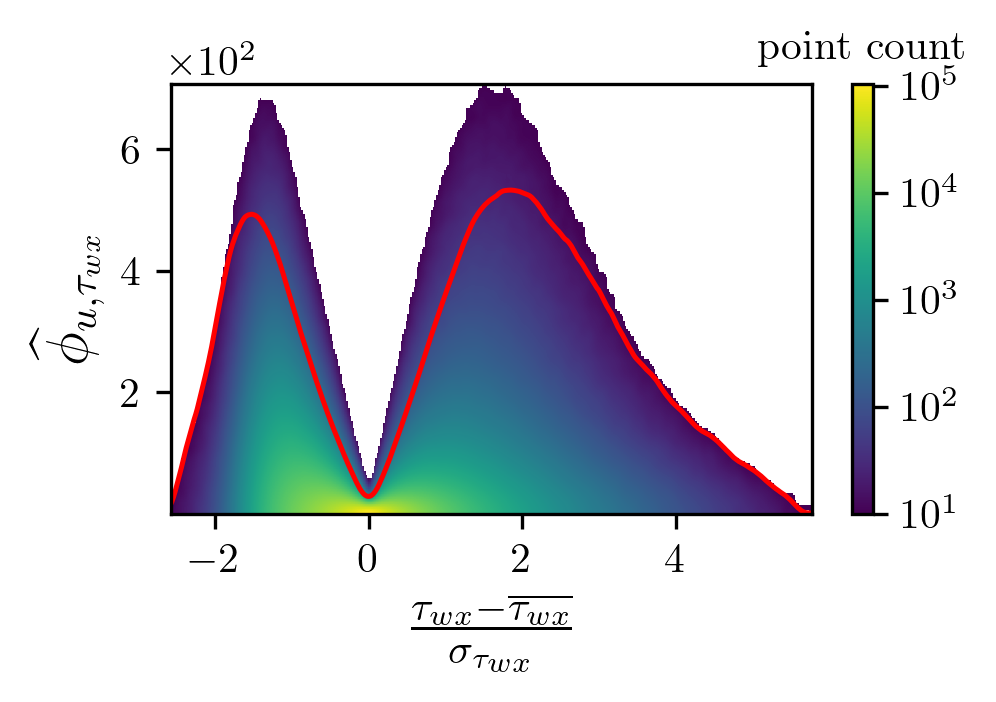}
\includegraphics[width=0.49\textwidth]{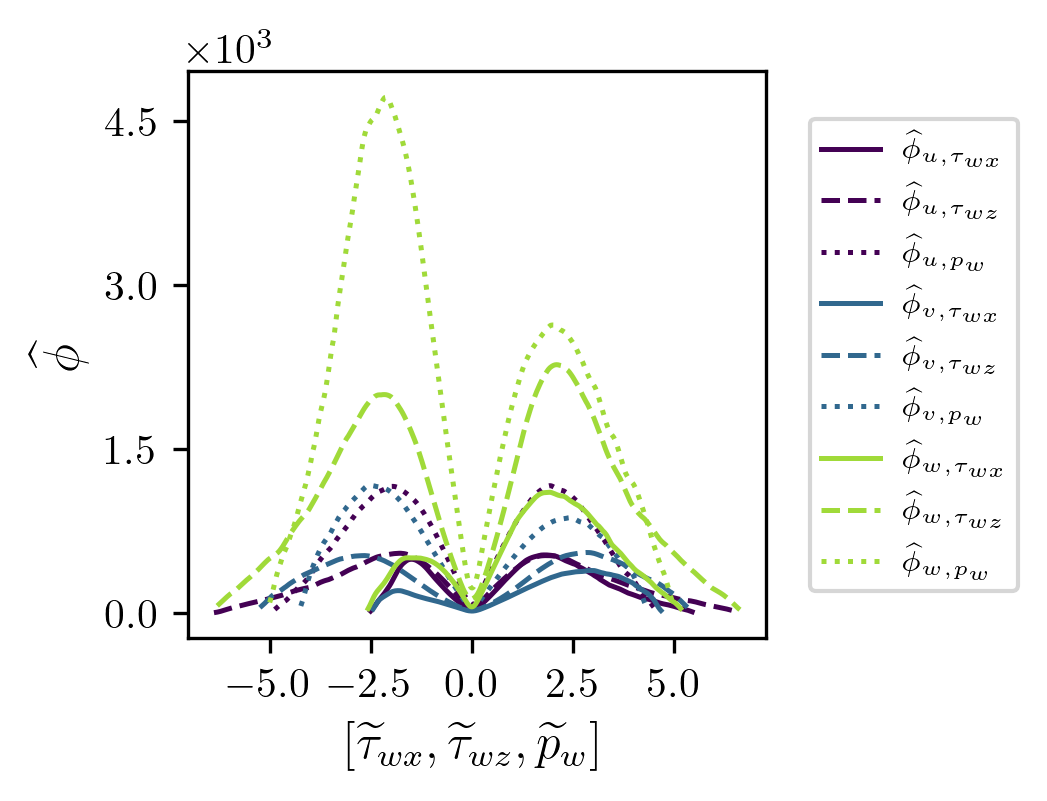}
\caption{Distribution of the SHAP values for the input feature values. The left figure shows the distribution of the $\tau_{wx}$ input SHAP values for the $u$ prediction collected over $\Delta t^+\approx 5000$. The red line indicates the threshold of the 99th percentile. The right figure presents the 99th percentile for all the combination input-output.}
\label{fig:distribution}
\end{figure}

From a statistical point of view, Figure \ref{fig:distribution} (left) shows the joint histogram distribution between standardized streamwise wall-shear stress and its importance scores $\phi_{u,\tau_{wx}}$ to predict the streamwise velocity fluctuation $u$. Note that the figure accounts for the correlation of every grid point for a total time $\Delta t^+\approx 5000$. The figure shows that a wide range of points are located near the mean value of the input, $\widetilde{\tau}_{wx}=0$. However, these grid points are related to low-importance scores. The higher importance is located from one to two standard deviations, while the domain areas containing peak values of the streamwise wall shear stress are less likely and do not present a high-importance content. This fact provides an important idea: the SHAP values do not directly correlate with wall shear stress values, similarly as observed by \citet{cremades2024} for the evaluation of the intense Reynolds stress structures~\citep{Lozano2012,Jimenez2018}.

%The SHAP values detect the high-importance regions of the model, which represents a trajectory of the turbulent flow. Therefore, we can infer that the SHAP values detect the regions of the flow that are more influential for the reconstruction of the fluctuations of the velocity. High-interest regions are those with a high absolute SHAP value.

%As previously stated, the high-intensity regions of the input fields are not the most influential of the field. Additionally, the distribution is slightly skewed, with the highest values of the streamwise shear stress located on its positive side. This also affects the peaks, being the peak of importance for positive shear stress obtained for higher absolute values than for the negative stress.

%The curves related to the streamwise velocity fluctuation present the highest SHAP values, as the turbulent structures are dragged in this direction by the pressure gradient. Moreover, the model predicts the velocity fluctuation at a wall-normal distance, $y^+=15$. This distance corresponds to the peak of the streamwise normal Reynolds stress~\citep{hoyas2008}, where, as previously stated, the streamwise streaks are more intense.

In addition, to easily compare all input-output pairs a threshold line is created. This line (red line of the left image in Figure \ref{fig:distribution}) represents the 99th percentile of the SHAP value. The domain below the line contains 99\% of the input scores for the respective field value. The 99th percentile line is used for comparing the different pairs of input-output combinations, Figure \ref{fig:distribution} (right). Concerning the inputs, the pressure is the most influential variable for the prediction of the velocity fluctuations, for a model in which both pressure and shear stress in the wall are known. Therefore, although for this wall-normal distance the wall-shear stress has been shown to be the most influential parameter in the reconstruction of the streamwise velocity~\citep{cuellar2024_2}, once the model knows both shear stress and pressure, an abrupt modification of the pressure might increment its error more than in the shear stress. This demonstrates that the estimator is more sensitive to missing or corrupted pressure information.

%This demonstrates that the flow is more sensitive to pressure variations.

Regarding the shear stress, both the streamwise and spanwise exhibit similar maximum values for the prediction of the streamwise velocity. Although this idea is non-intuitive, as the streamwise shear could be expected to produce higher effects on the streamwise velocity than the spanwise, the model is trained to predict the velocity based on the combined effect of the wall-pressure and the wall-shear stress. This fact results in higher errors when the correlation between the input variables is broken; for this reason, the pressure becomes the most influential and reduces the effect of the streamwise wall-shear stress.

Completely different effects are observed in the case of the spanwise velocity. As could be expected, the spanwise velocity is more affected by the spanwise shear, which forces the motion transversely to the main forcing of the channel, than by the streamwise shear, which is expected to force the motion in the same direction as the pressure gradient. Finally, similar trends are observed for the wall-normal velocity, which is mainly affected by the variations on the pressure and then in the spanwise shear stress.

%Finally, the wall-normal velocity is the less sensitive component of the velocity fluctuation, as the SHAP values related to its evolution, which are a simplification of the MSE of the reconstruction of the velocity field, are one order of magnitude lower than for the other components.

The distribution of Figure \ref{fig:distribution} (right) shows that for all the input-output pairs, the higher importance is not obtained for the most intense input regions. The agreement between the 20\% most important and the 20\% most intense input regions remains below 60\% of their surface. In addition, the most important regions neither match the intense output locations, being the coincidence lower than 45\%. Therefore, those regions affecting the predictions the most have been demonstrated to be strongly related to the model instead of the intense input and output regions. However, a new question is raised: How much does each percentile of importance condition the estimation?

\begin{figure}
\includegraphics[width=0.49\textwidth]{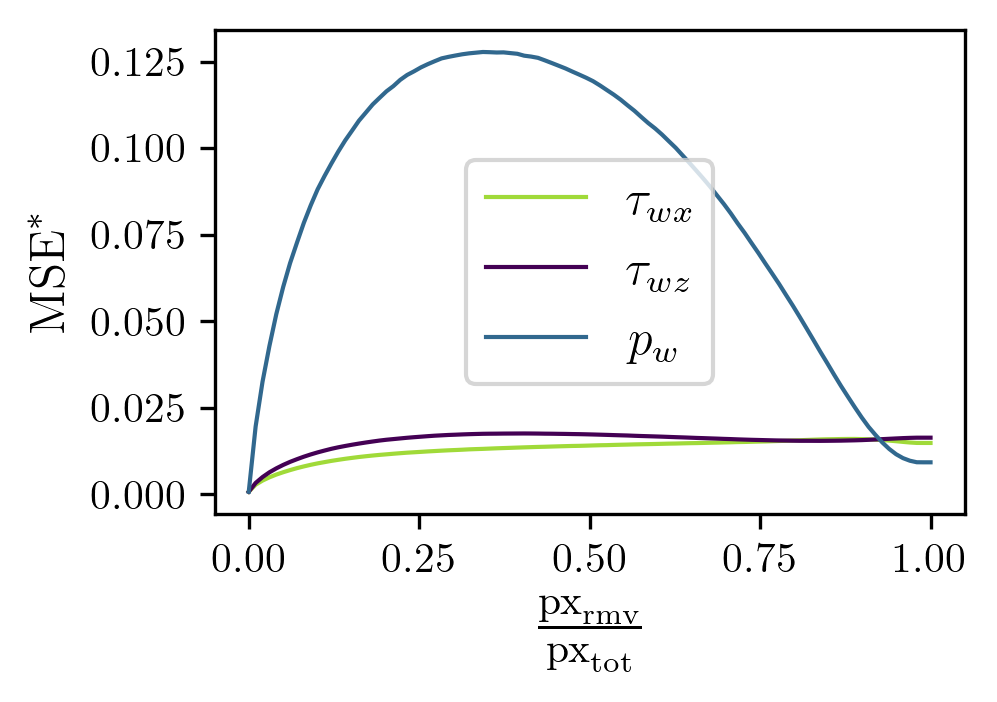}
\includegraphics[width=0.49\textwidth]{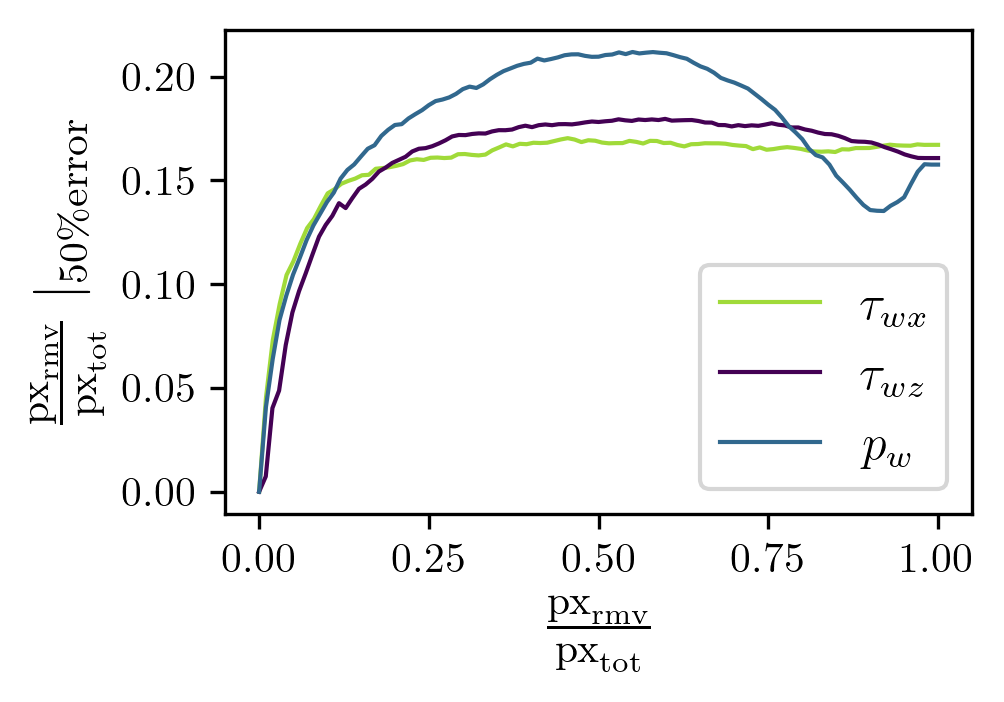}
\caption{Mean Squared Error of velocity fluctuations as a function of the fraction of pixels removed from the input (left) and fraction of pixels of the prediction containing 50\% of the total MSE score for the streamwise velocity fluctuation (right). The fraction of grid-point removal relative to their total number is presented by $\text{px}_\text{rmv}/\text{px}_\text{tot}$, and the fraction of grid-point removal for half of the error is $\left.\text{px}_\text{rmv}/\text{px}_\text{tot}\right|_{50\%\text{error}}$.}
\label{fig:MSE_all_y15}
\end{figure}

To answer the previous question, a certain fraction containing the highest nth percentile of grid points is removed from the input field and set to their mean value. In other words, the pixels from the input fields are gradually removed according to their importance, evaluating the accuracy of the predictions progressively. The accuracy is evaluated by measuring the mean squared error (MSE) of the predicted velocity fluctuations. The evaluation of the MSE as a function of the percentage of top-grid points removed is presented in Figure \ref{fig:MSE_all_y15}. The normalized MSE, $\text{MSE}^*$ of the figure is calculated as follows:

\begin{equation}
\text{MSE}^*= \frac{1}{3}\left(\frac{\text{MSE}_u}{u_{\rm{max}}}+\frac{\text{MSE}_v}{v_{\rm{max}}}+\frac{\text{MSE}_w}{w_{\rm{max}}}\right)
\end{equation}

\begin{figure}
\includegraphics[width=0.49\textwidth]{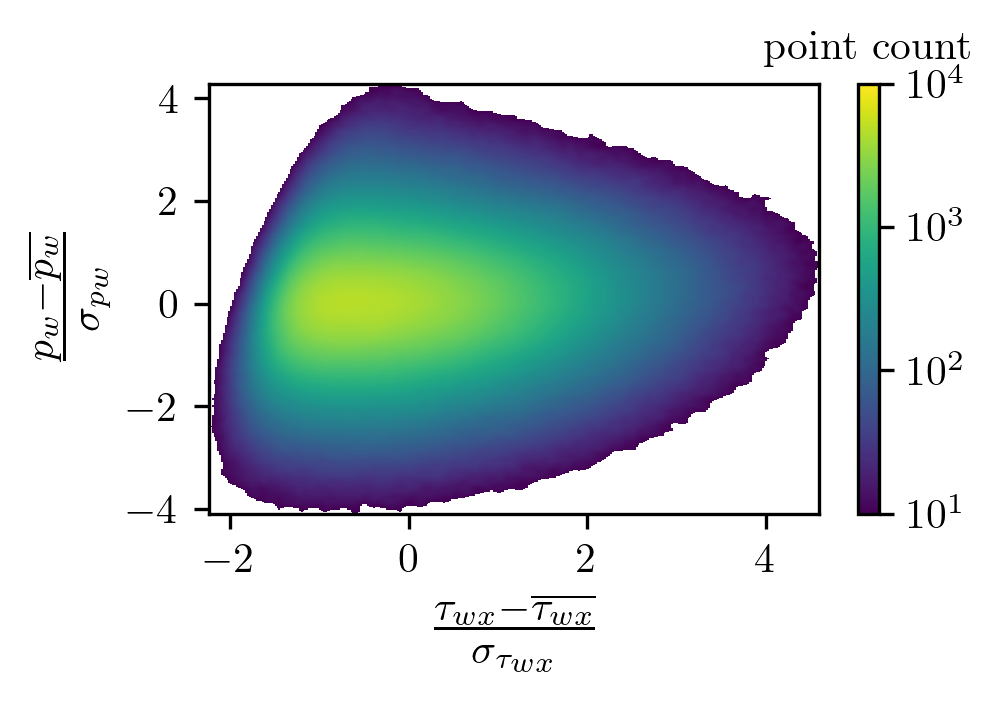}
\includegraphics[width=0.49\textwidth]{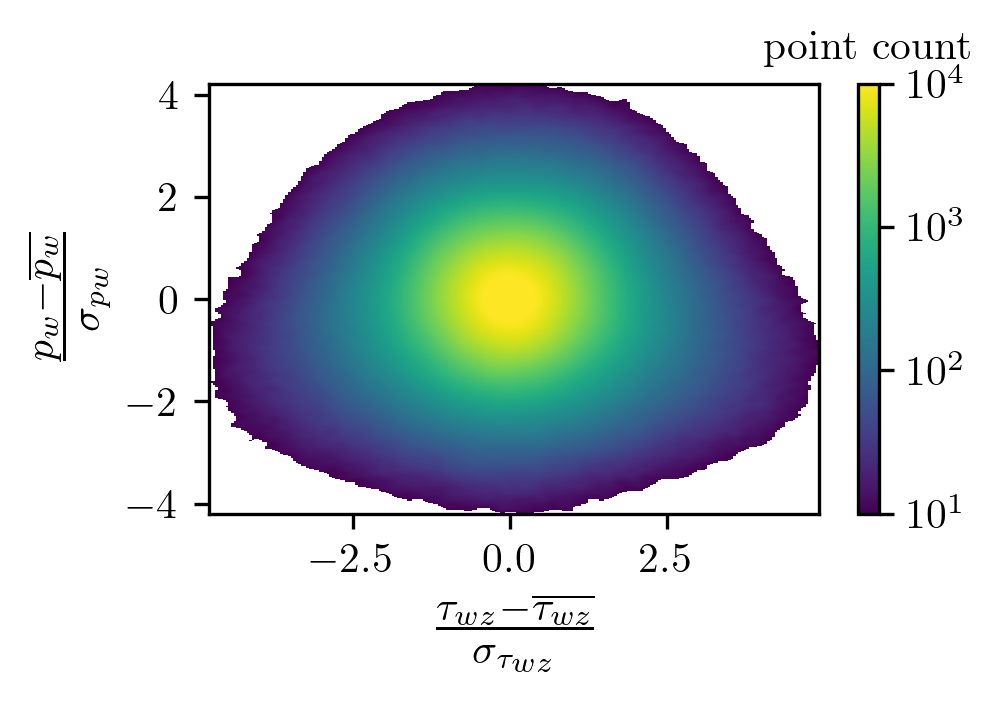}
\caption{Probability density function} of the wall-pressure and wall-shear stress in the streamwise direction (left) and the spanwise direction (right).
\label{fig:distributionptau}
\end{figure}

The results of Figure \ref{fig:MSE_all_y15} (left) correspond to different inputs being modified separately. As expected after the analysis presented in Figure \ref{fig:distribution}, the most influential input field is the wall pressure, while wall shear stress in both streamwise and spanwise directions is less important. The results for the pressure show that a fast increase in the error is produced when the initial percentiles of important grid points are removed. A maximum error of the MSE of 12\% of the maximum value of the original velocity fields is obtained when approximately 30\% of the grid points are removed. After this point, the input field becomes non-informative and therefore, the error starts to decrease. The decrease is not observed in the case of the wall shear stress due to the high influence of the pressure in the predictions. This effect is produced by the strong correlation between the pressure and shear stress fluctuations. When the pressure fluctuation is affected, its variation results in an increase of error of the model, which is unable to detect the combined patterns of both magnitudes. Figure \ref{fig:distribution} showed that the maximum importance was obtained for intense values of the standardized wall-pressure and wall-shear stress between $1.5$ and $2.5$. As can be observed in Figure \ref{fig:distributionptau}, a non-negligible percentage of the intense-shear stress regions, $\left\vert\widetilde{\tau}_{wx}\right\vert\ge1.5$ and $\left\vert\widetilde{\tau}_{wz}\right\vert\ge1.5$, correlate with intense wall-pressure regions, $\left\vert\widetilde{p}_{w}\right\vert\ge1.5$. For this reason, although the streamwise velocity fluctuation at a wall-normal distance $y^+=15$ correlates mostly with the streamwise shear stress, modifying the pressure associated with this region modifies the pattern detected by the deep-learning model, increasing the error. However, when the percentage of removed pixels is increased, the pressure field becomes non-informative and the model only relies on the values of the shear stress. The error produced by the streamwise and spanwise shear stress reaches a plateau when approximately 20\% of pixels have been removed and then it remains around 1.5\%. This plateau is produced due to the high influence of the pressure in the prediction of the flow, as when the shear stress becomes non-informative, the model relies on the pressure.

\begin{figure}
\centering
\includegraphics[width=\textwidth]{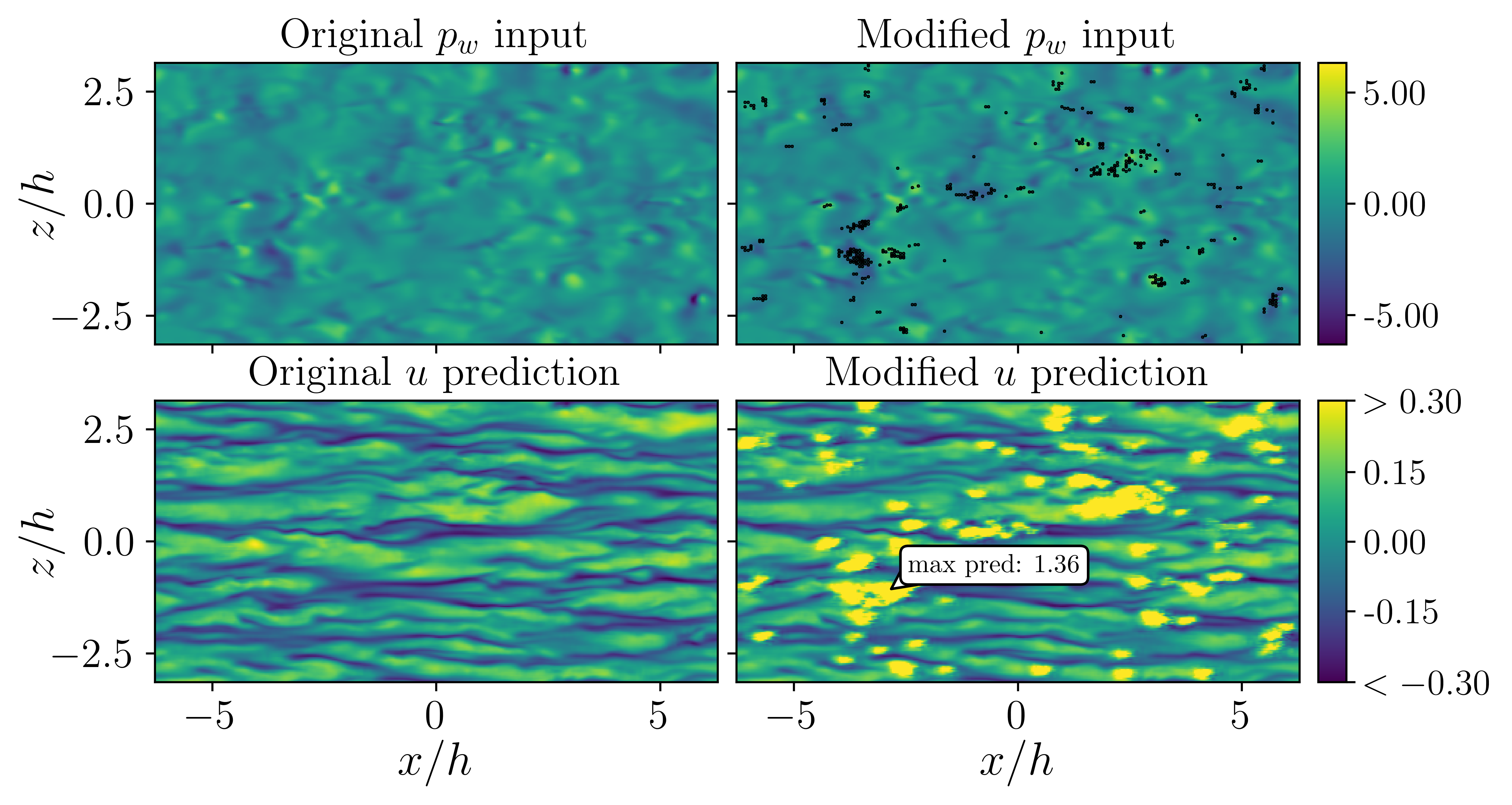}
\caption{Comparison between original (left) and modified (right) wall pressure with top 1\% absolute SHAP values removed from the predictions of the streamwise velocity. The top figures show the wall pressure field and the bottom images the predicted streamwise velocity. The black dots in the upper right image indicate the input pixels replaced by the non-informative background value, while the modified streamwise velocity prediction is capped at the original prediction maximum values to indicate the changes in prediction.}
\label{fig:comp_p}
\end{figure}

The high impact of the top-ranked grid points is presented in Figure \ref{fig:comp_p}. In the figure, the original wall pressure field is presented (top-left image). This field is modified by removing the top 1\% grid points (black contours in the top-right image) and setting them to their reference value. The bottom images show the original and modified prediction. The figure evidences how a small percentage of pixels, 1\%, modify a much larger region (bottom-right image), which overflows the predictions, generating a velocity value much higher than the original. The analysis of the modified regions of the flow matches the effect of the receptive field. The CNN presented in \citet{Guastoni2020} has a receptive field of $15\times 15$, meaning that each input value can affect the 15 adjacent values, as can be observed in the bottom-right image of Figure \ref{fig:comp_p}.

Then, in Figure \ref{fig:MSE_all_y15} (right), the fraction of grid points of the output contributing to half of the error value is visualized against the fraction of grid points removed from the input. The results show that initially when the first decile of input grid points is removed, half of the total error is concentrated in a small area, below 10\% of the total output. However, as the fraction of the removed input grid points is increased, the error is propagated up to 15\% of the output field in the case of the shear stress and 20\% in the case of the pressure. From this point in advance, half of the error remains concentrated in around 20\% of the output domain.

\begin{figure}
\centering
\makebox[0.49\textwidth]{\includegraphics[height=4.2cm]{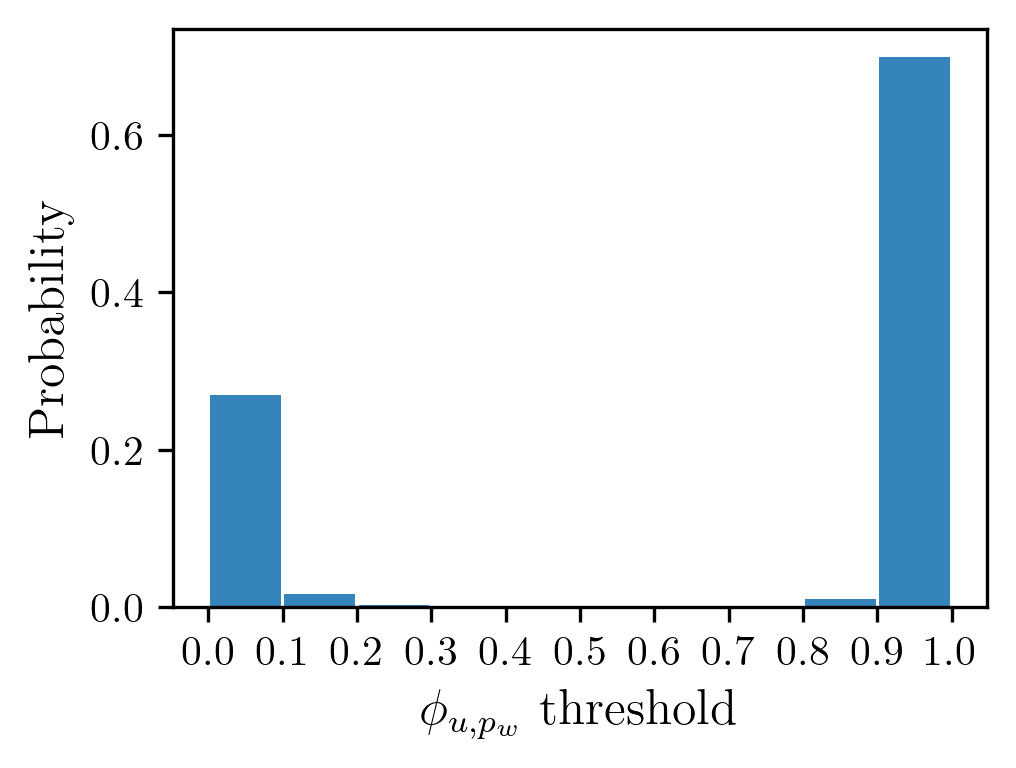} }
\includegraphics[width=0.49\textwidth]{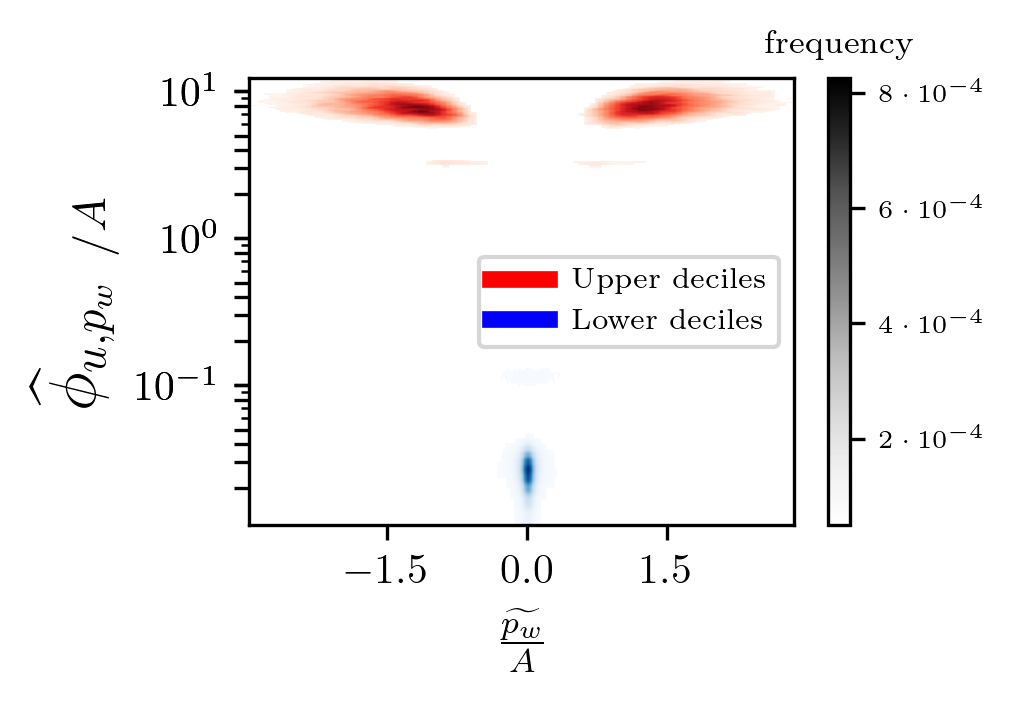}
\includegraphics[width=0.49\textwidth]{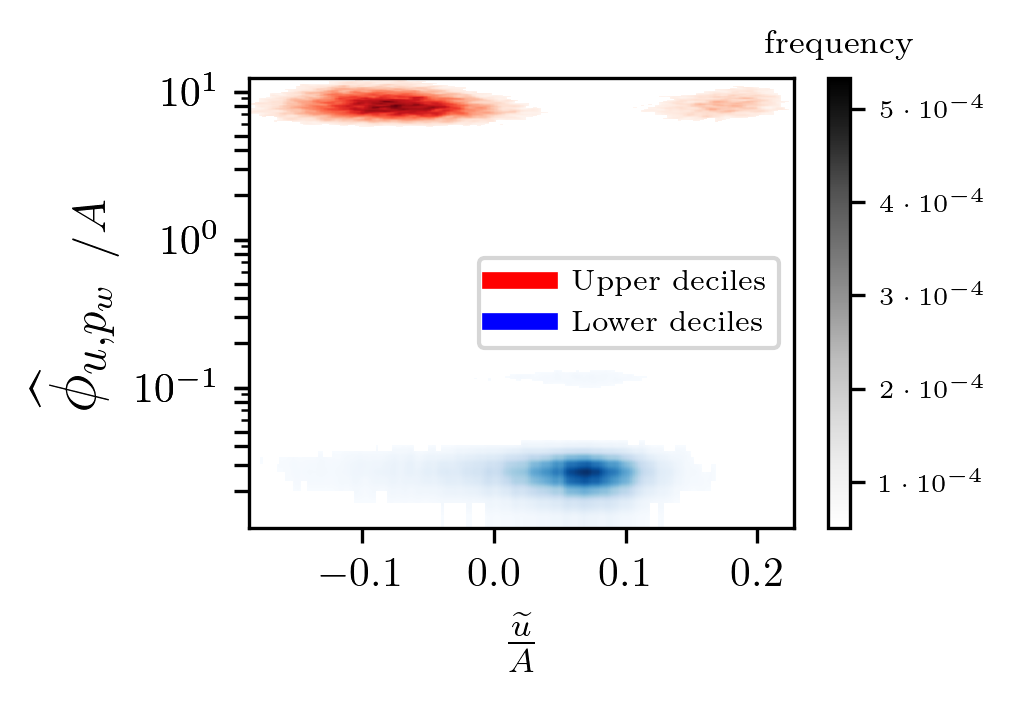}
\includegraphics[width=0.49\textwidth]{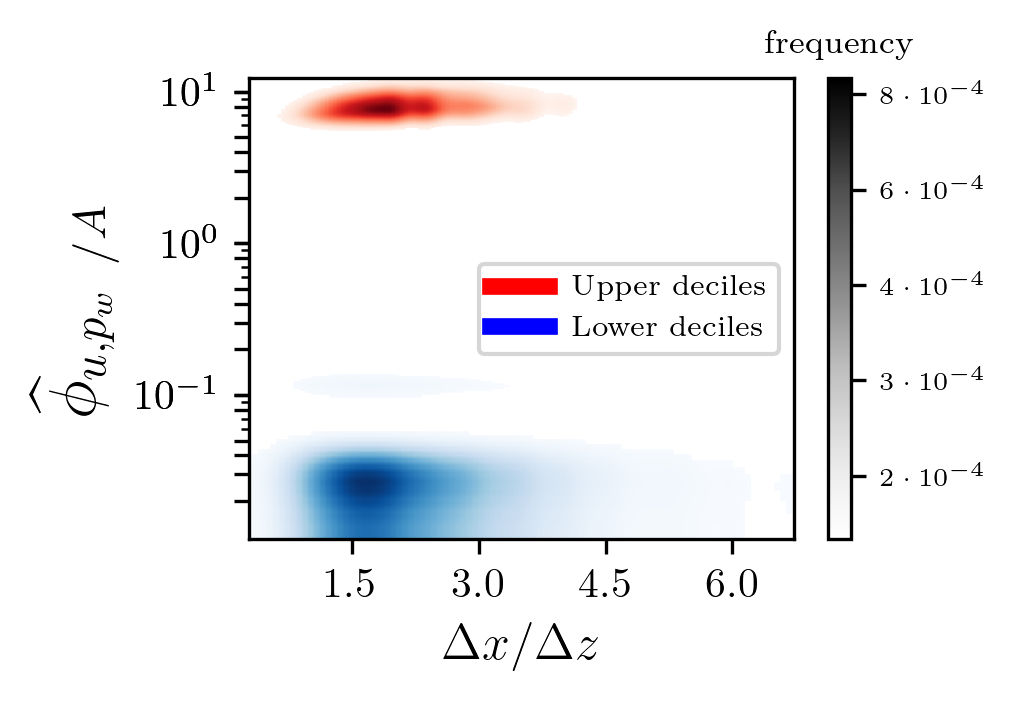}
\includegraphics[width=0.49\textwidth]{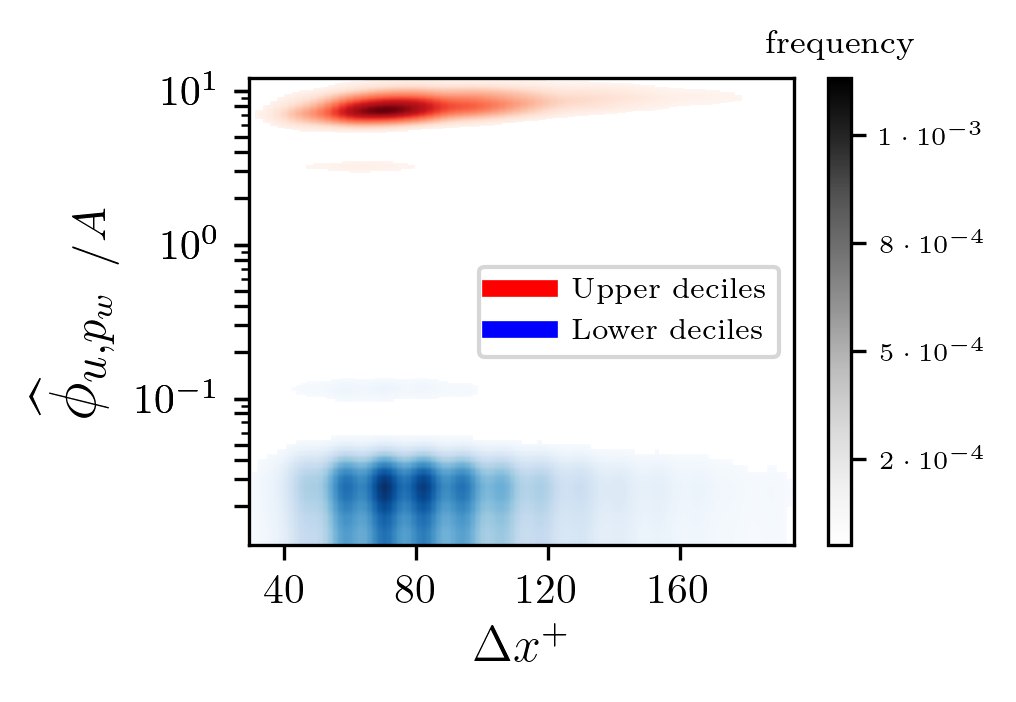}
\includegraphics[width=0.49\textwidth]{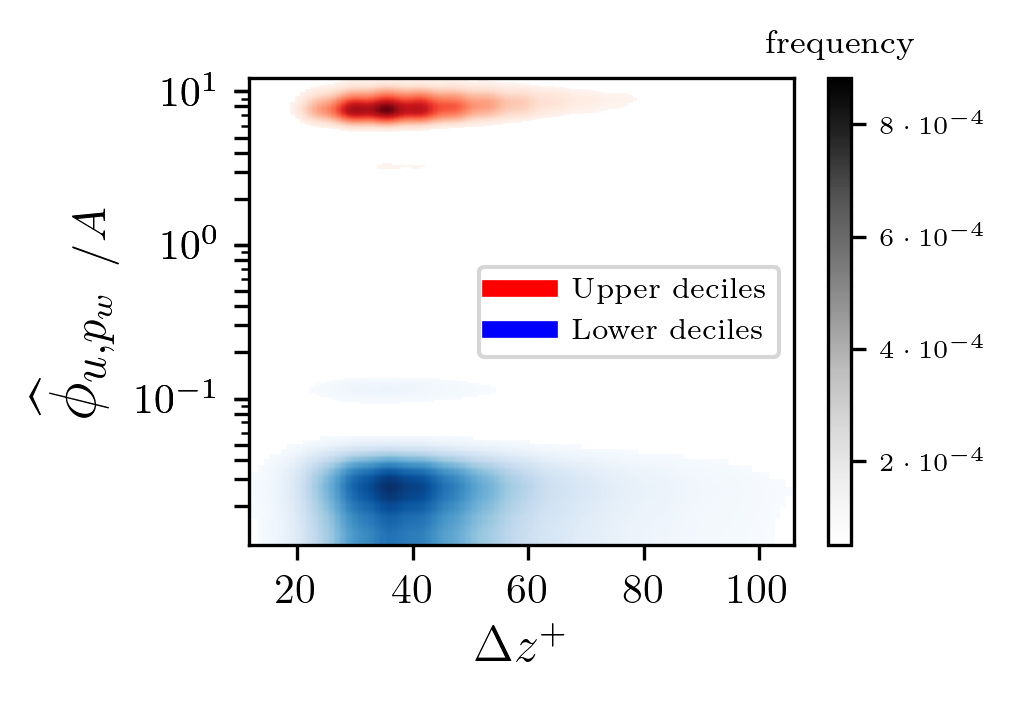}
\caption{Probability of the most important clusters for predicting the streamwise velocity from the wall pressure. Probability of the structures belonging to each percentile of importance (top-left), and joint probability of their averaged physical properties and importance: pressure (top-right), streamwise velocity (middle-left), aspect ratio (middle-right), streamwise length (bottom-left) and spanwise width (bottom-right).}
\label{fig:bincount_15}
\end{figure}

To understand the input structures that are conditioning the prediction of the model, the domain is segmented in deciles. Then, the grid points corresponding to each decile segmentation are joined into structures, filtering those with an area in viscous units smaller than $S^+ = 30^2$~\citep{Lozano2012}. The probability of the filtered structures belonging to each importance decile is calculated over 1000 snapshots, $\Delta t^+\approx 5000$, as described in Figure \ref{fig:bincount_15} (top-left). The figure evidences that only the structures of the first and the last deciles are significant, the rest are too small to pass the filter as they are scattered around and are not big enough to form clusters. Finally, the averaged physical properties of the formed clusters are analyzed against their impact on the prediction to understand the common properties of the high and low-importance regions.

The analysis of the wall pressure averaged over the area of the most important and the least important deciles of structures are presented in Figure \ref{fig:bincount_15} (top-right). As previously discussed, pressure is the input variable with a higher impact on the predictions at a wall distance $y^+=15$. The figure shows symmetrical distributions for the high-impact structures (red), which are distributed from one to two standard deviations from the mean pressure, showing equivalent impact in the solution for positive and negative fluctuations of the pressure. On the other hand, a low-importance region is detected for averaged pressures. Then, the values of the averaged streamwise velocity predicted over this region are analyzed. The results evidenced that the high-importance regions on the wall are mostly located below low streamwise velocity areas, low-velocity streaks, or ejection-like regions. On the other hand, the lower-importance regions are slightly skewed to positive-velocity values. This idea is consistent with the results of \citet{cremades2024}, in which the ejection-like regions were demonstrated to be the most influential for the evolution of the flow at low friction Reynolds numbers. In addition, note the small values of the deviation from the mean of the averaged velocities along the structure area. This fact shows that the high-importance regions contain high and low-velocity areas and also connect with the results of \citet{cremades2024} which evidenced that the most important areas are not those with the highest Reynolds stress. Concerning the shape and size of the structures, the high-intensity structures are slightly elongated in the streamwise direction. This elongation is a result of the anisotropy of the turbulence in the channel. The flow forces the structures to move in the streamwise direction creating elongated patterns. In addition, the size of the structures, in both streamwise and spanwise directions, is evaluated to provide an objective criterion for placing the sensors. Most of the important structures measure around $80\times40$ wall units. In the case of the study, the grid point separation is $\Delta x^+\approx 12$ for the streamwise direction and $\Delta z^+\approx 6$ for the spanwise, which corresponds to approximately 6 measurements in each direction of the cluster. Note that this is a crucial factor for industrial applications, as undersampling needs to be avoided. For instance, \citet{haakansson2013} required 20\% of a particle streak size to measure to avoid altering its width. \citet{liu1991} reported a spacing for experimental particle image velocimety of $7.5$ and $6$ wall units for the streamwise and spanwise directions respectively. According to \citet{zhe2005}, the shear sensors have a resolution of the order of 100\si{\micro\meter}, which corresponds to approximately $1.7$ viscous units for the conditions of the wind-tunnel experiments of \cite{ganapathisubramani2005}. However, these sizes present a challenge for applications in commercial flights because in these conditions the averaged unitary viscous length corresponds to approximately 5\si{\micro\meter}. Therefore, for the friction Reynolds number corresponding to a commercial aircraft flight, as the most important structures exhibit a size between $20$ and $120$ wall units, the equivalent size for a commercial aircraft will be in the range from $100$ to $600$\si{\micro\meter}, meaning that with the previous sensor size only 1 to 6 points inside the cluster could be measured. For this reason, and according with the experience reported by \citet{haakansson2013}, the rule of measuring at least every $20$\% of the structure size might not be satisfied for some structures with the current technology. Misleading measurements might be taken in the case of a commercial flight, and thus, further research needs to be conducted in order to quantify the sensitivity of the reconstruction of the flow when coarser grids are used over the wall.

\begin{figure}
\centering
\includegraphics[width=0.49\textwidth]{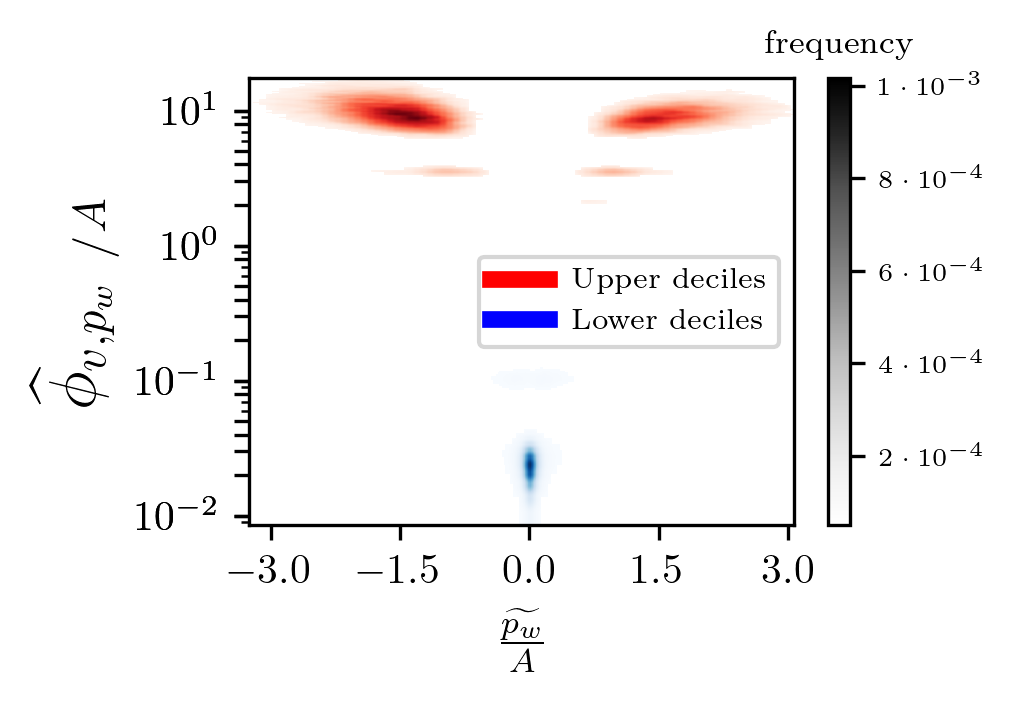}
\includegraphics[width=0.49\textwidth]{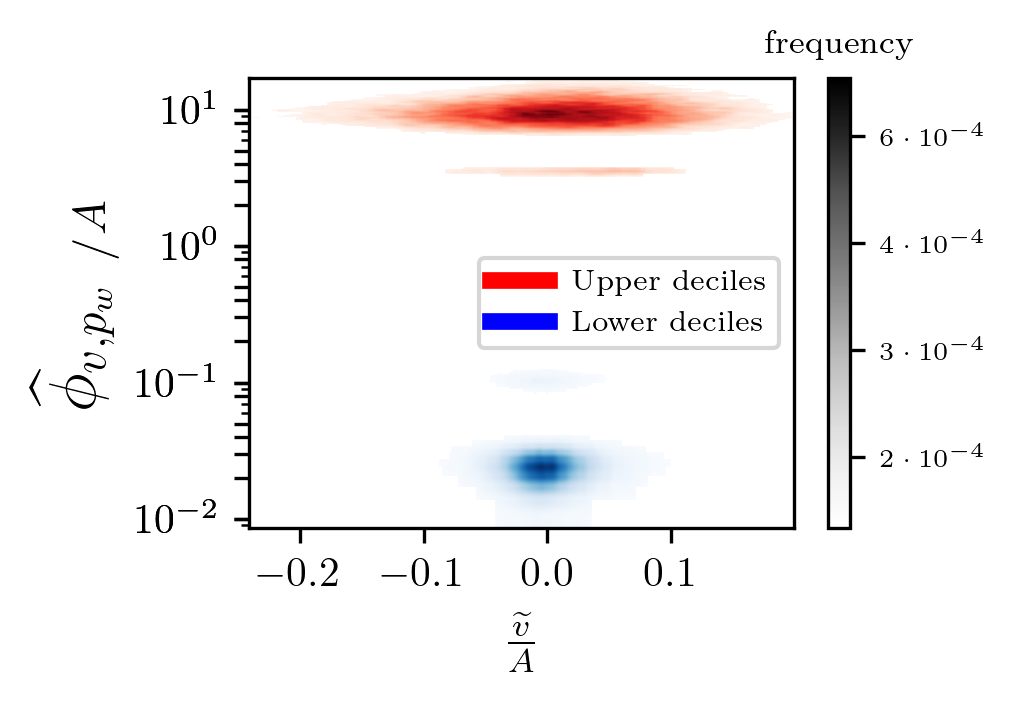}
\includegraphics[width=0.49\textwidth]{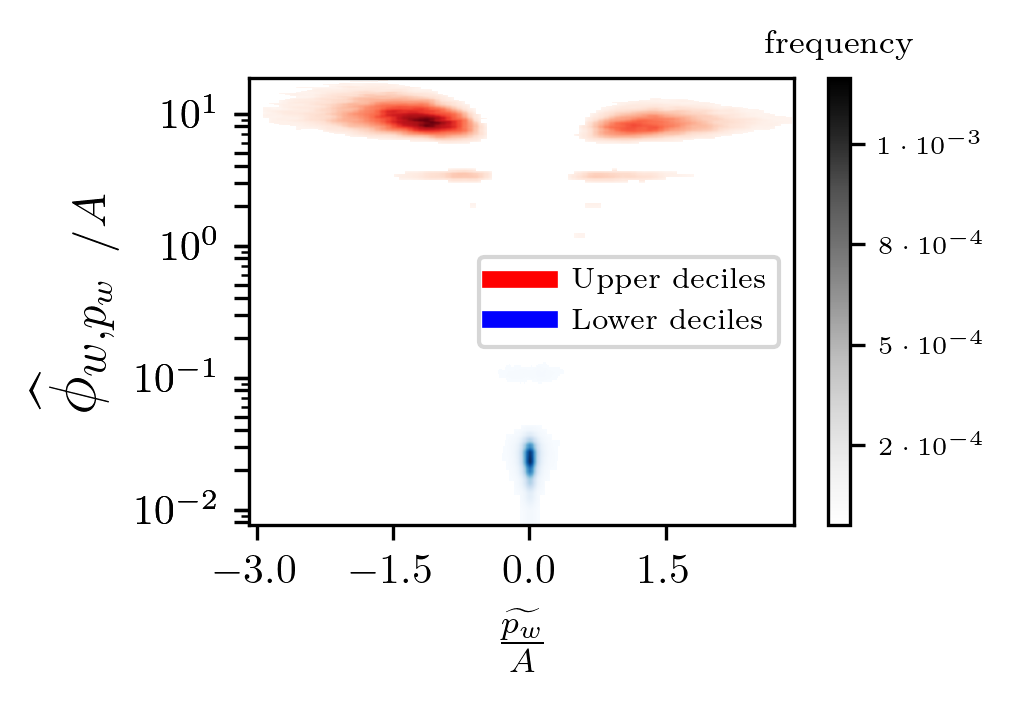}
\includegraphics[width=0.49\textwidth]{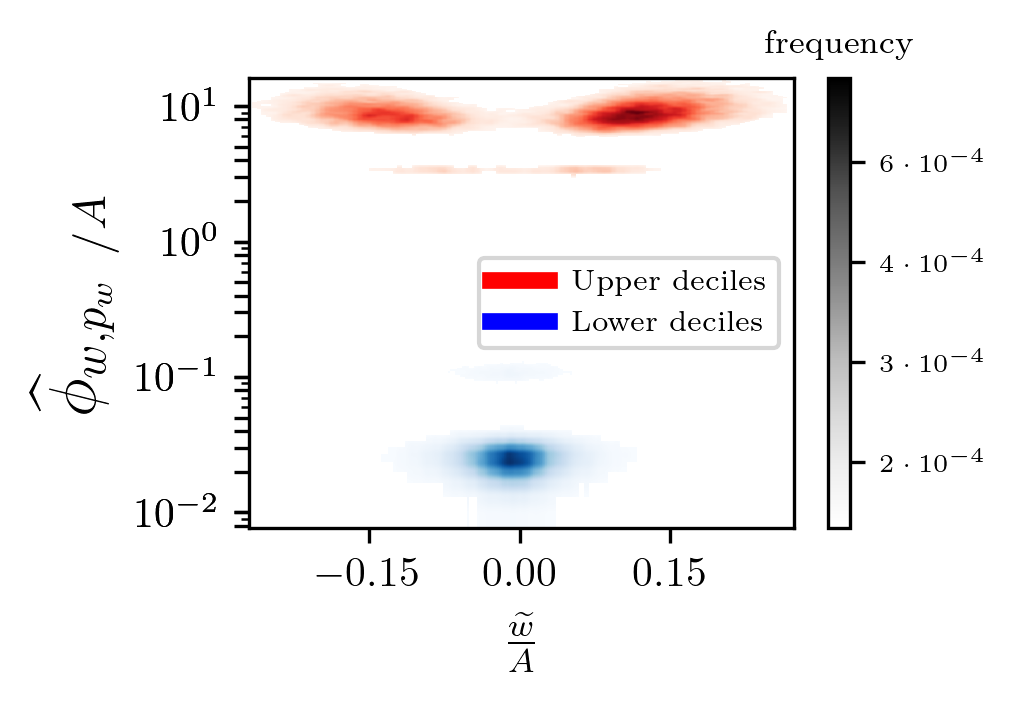}
\caption{Probability of the importance clusters for predicting the wall-normal and spanwise velocity from the wall pressure. Joint probability of the averaged physical properties of the structures, averaged pressure (left) and average velocity fluctuation (right), and the importance of the pressure for the prediction of wall-normal (top) and spanwise (bottom) velocity.}
\label{fig:phyprop_2}
\end{figure}

\begin{figure}
\centering
\includegraphics[width=0.49\textwidth]{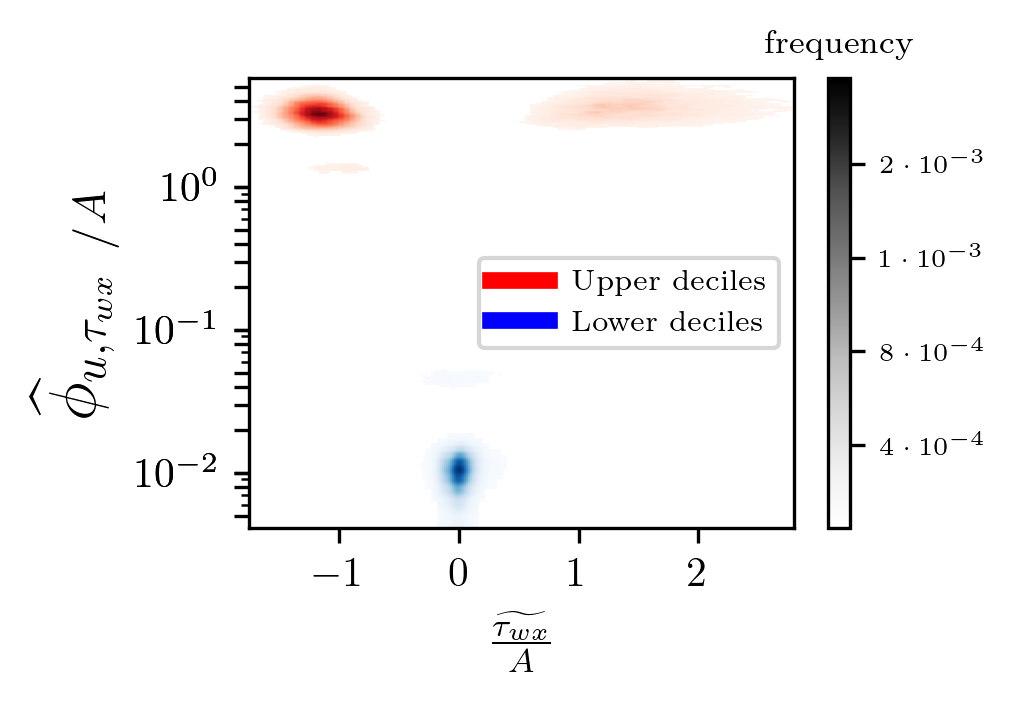}
\includegraphics[width=0.49\textwidth]{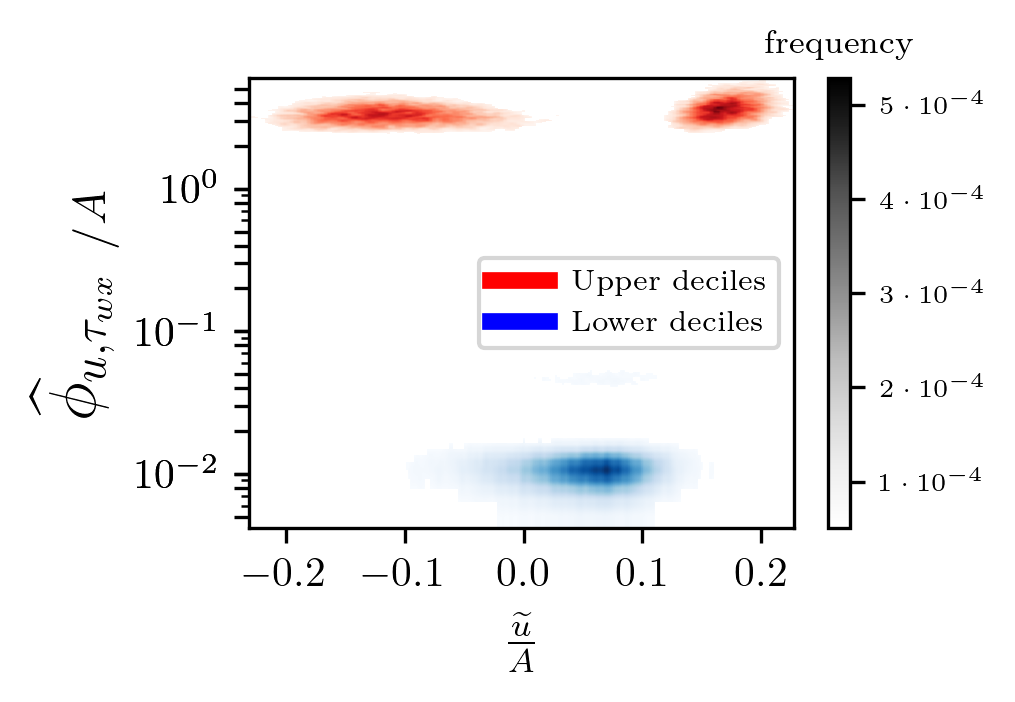}
\includegraphics[width=0.49\textwidth]{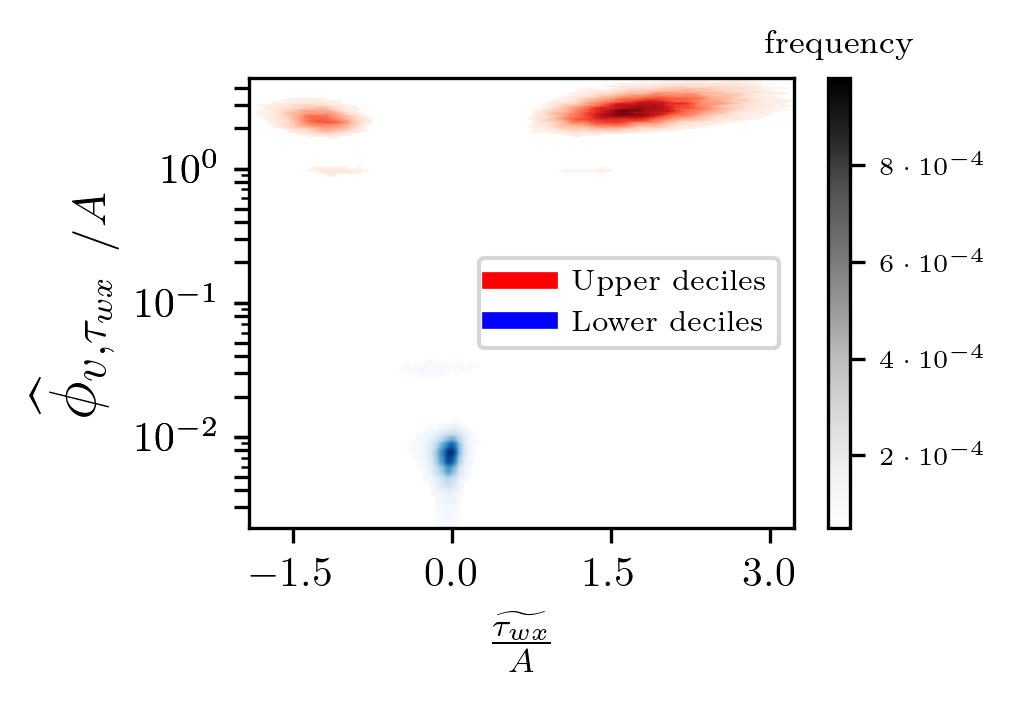}
\includegraphics[width=0.49\textwidth]{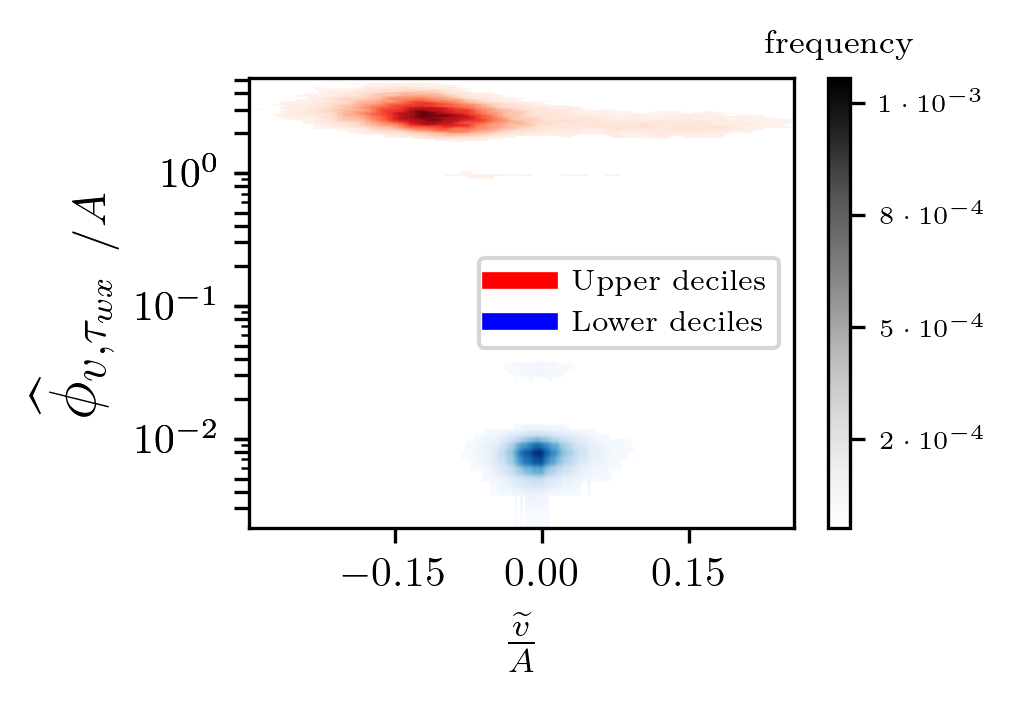}
\includegraphics[width=0.49\textwidth]{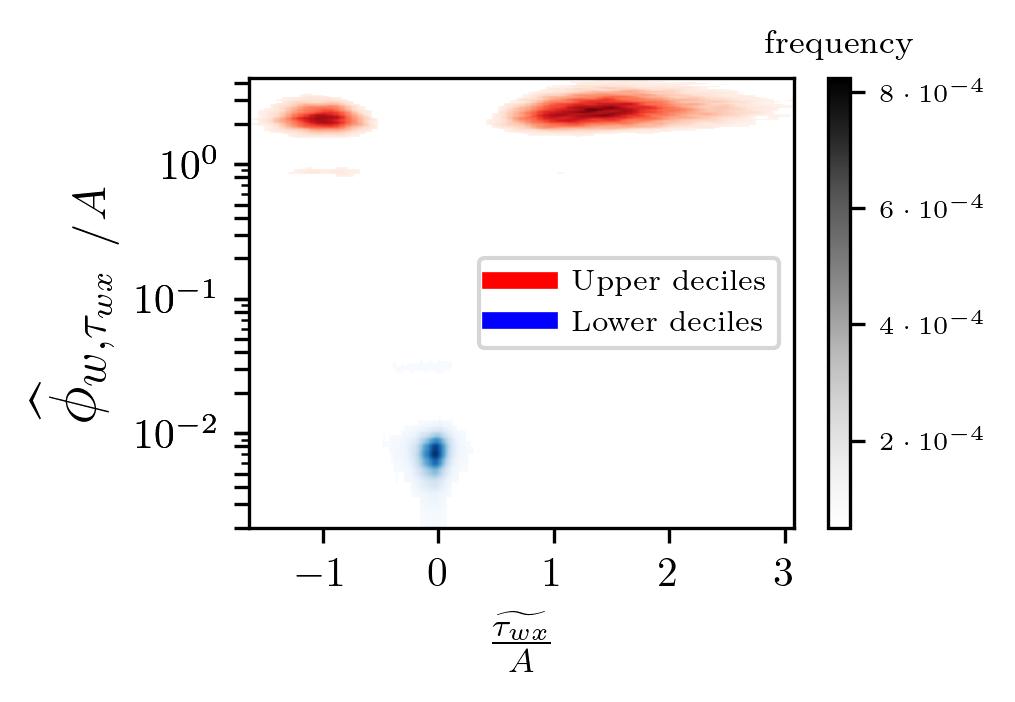}
\includegraphics[width=0.49\textwidth]{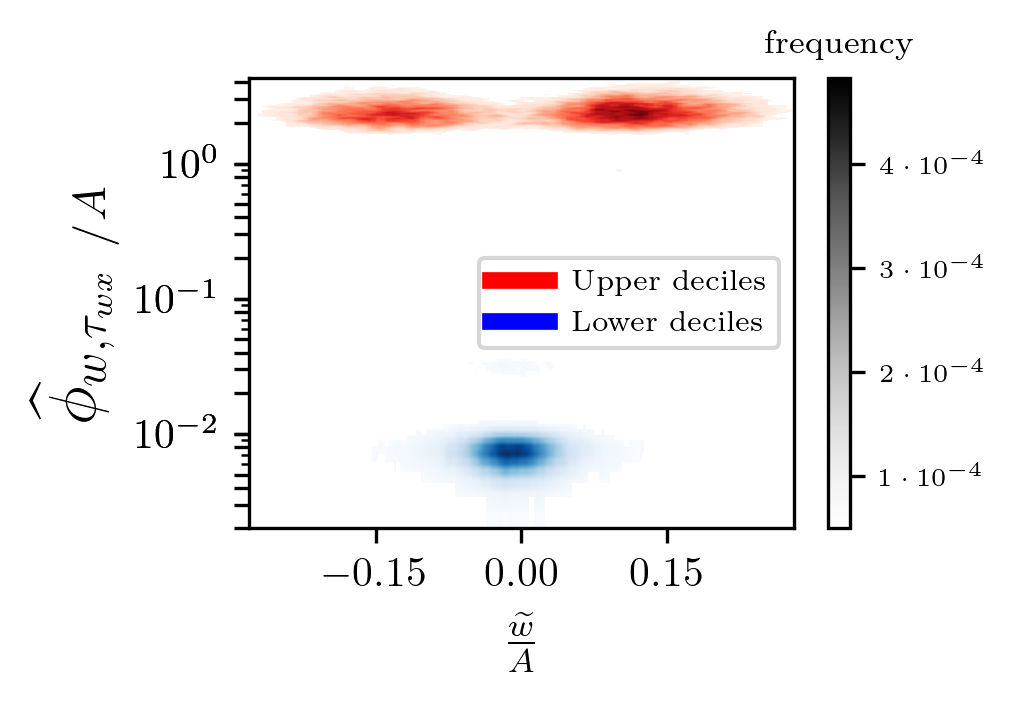}
\caption{Probability of the importance clusters for predicting the streamwise, wall-normal and spanwise velocity from the streamwise shear stress. Joint probability of the averaged physical properties of the structures, averaged streamwise shear stress (left) and average velocity fluctuation (right), and the importance of the pressure for the prediction of streamwise (top), wall-normal (middle) and spanwise (bottom) velocity.}
\label{fig:phyprop_0}
\end{figure}

\begin{figure}
\centering
\includegraphics[width=0.49\textwidth]{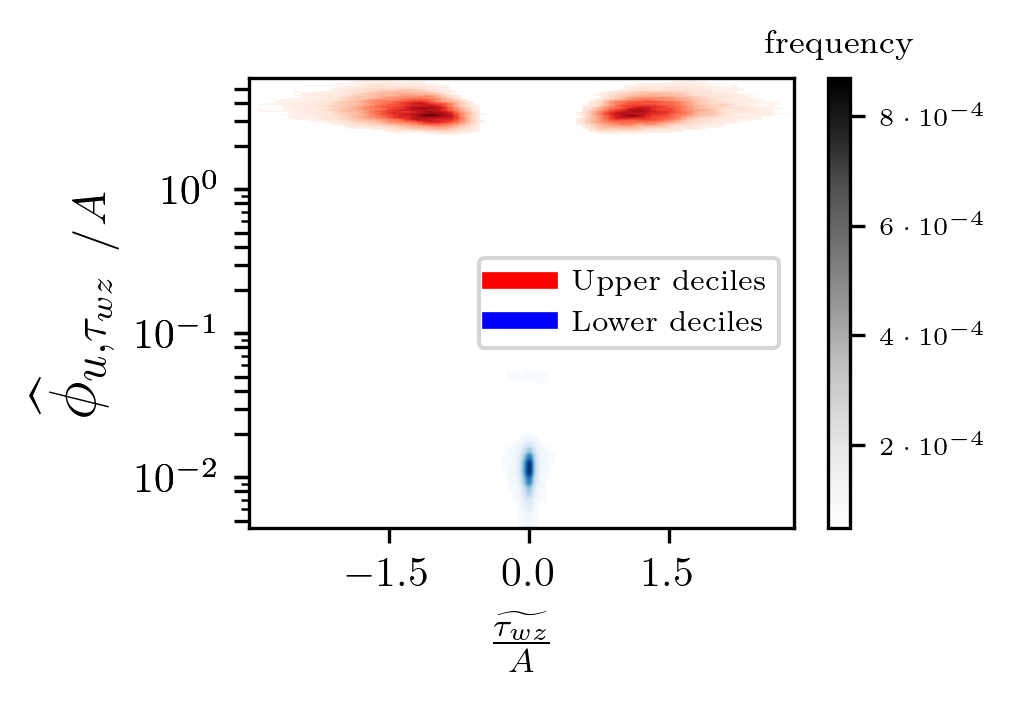}
\includegraphics[width=0.49\textwidth]{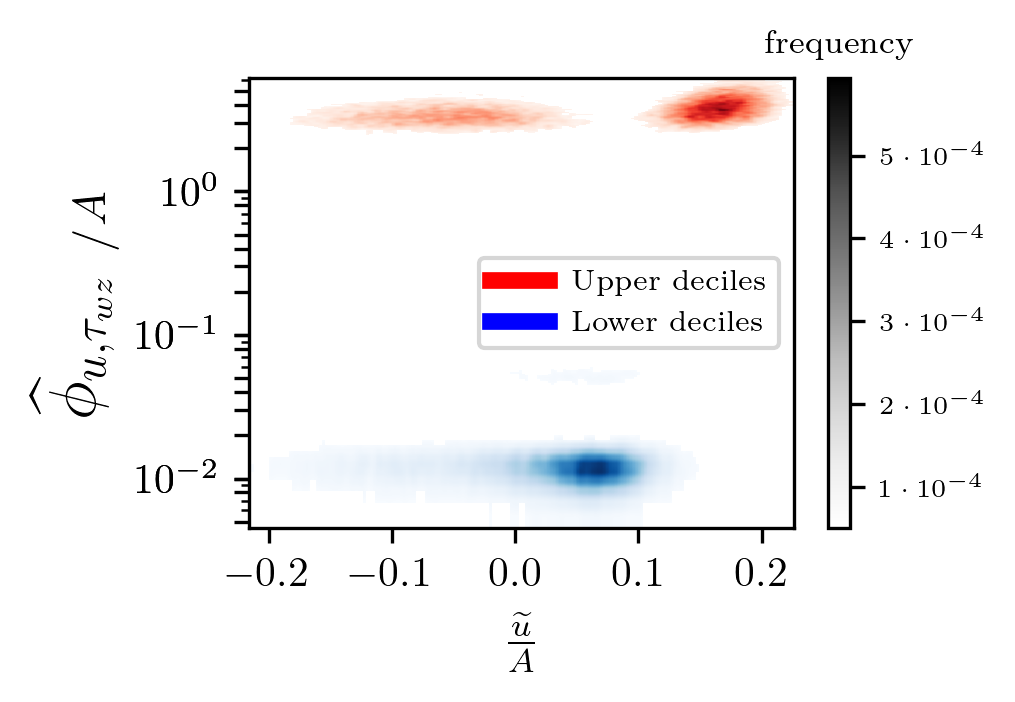}
\includegraphics[width=0.49\textwidth]{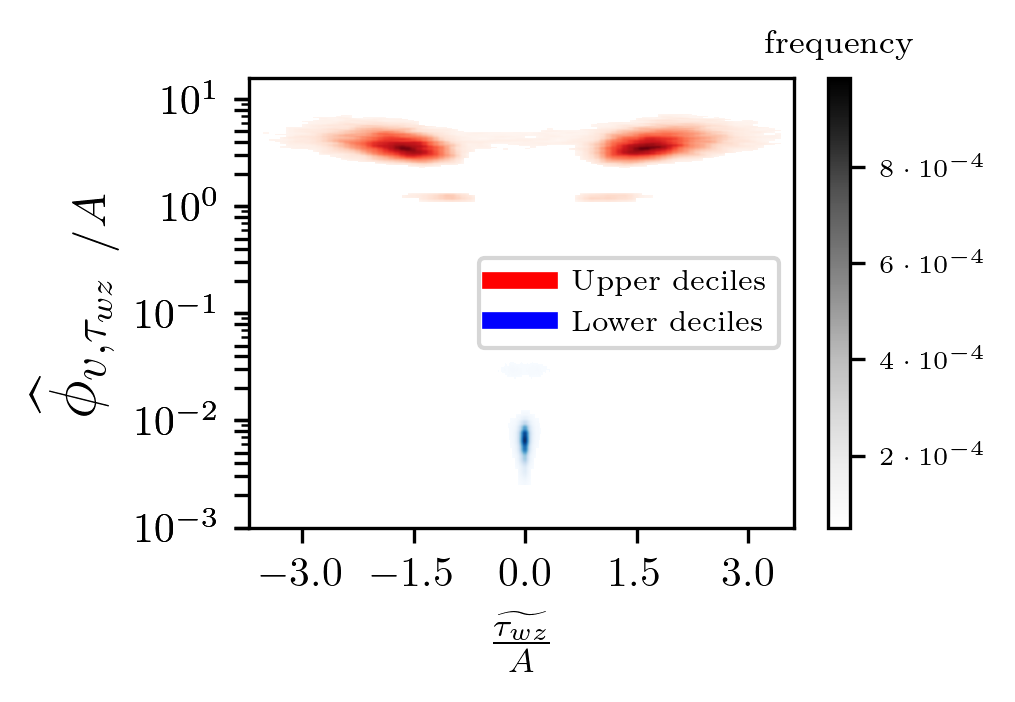}
\includegraphics[width=0.49\textwidth]{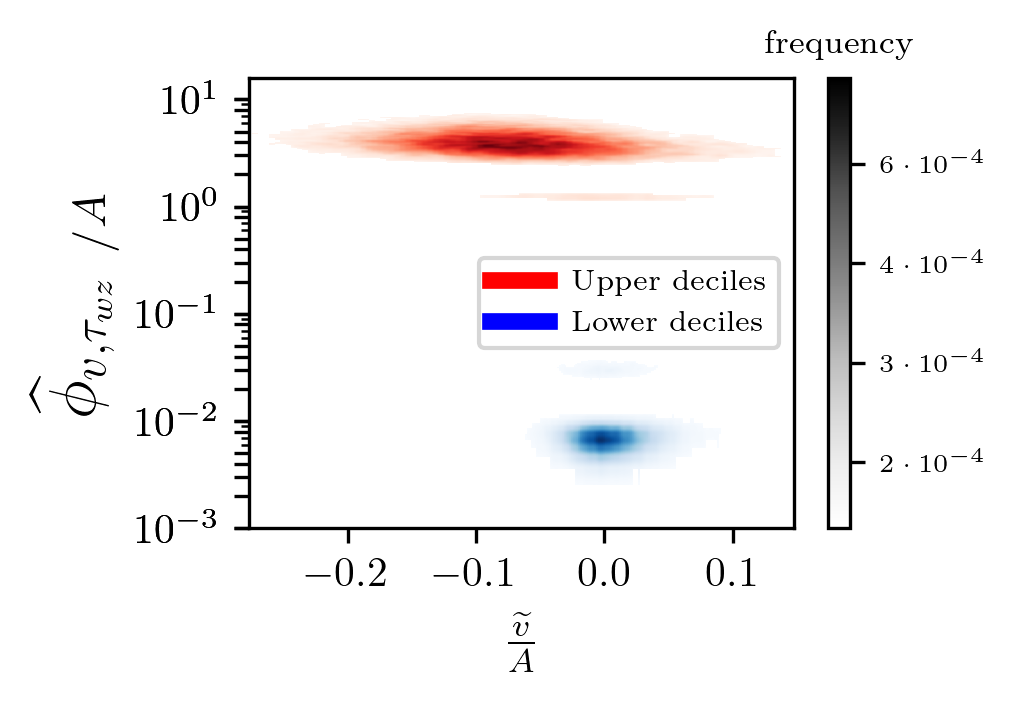}
\includegraphics[width=0.49\textwidth]{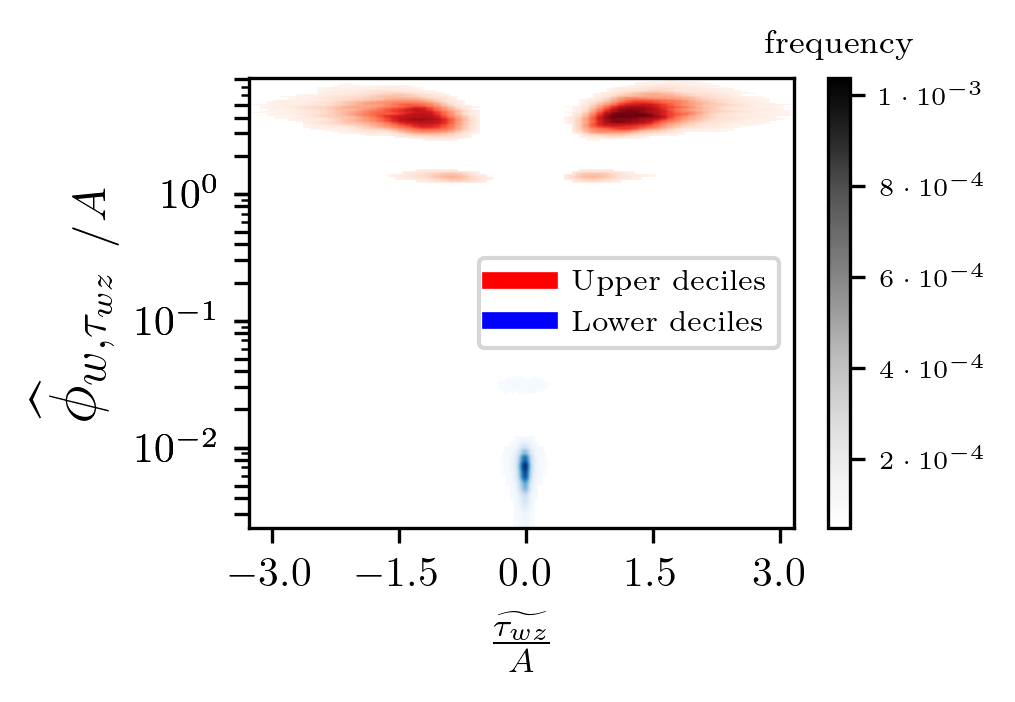}
\includegraphics[width=0.49\textwidth]{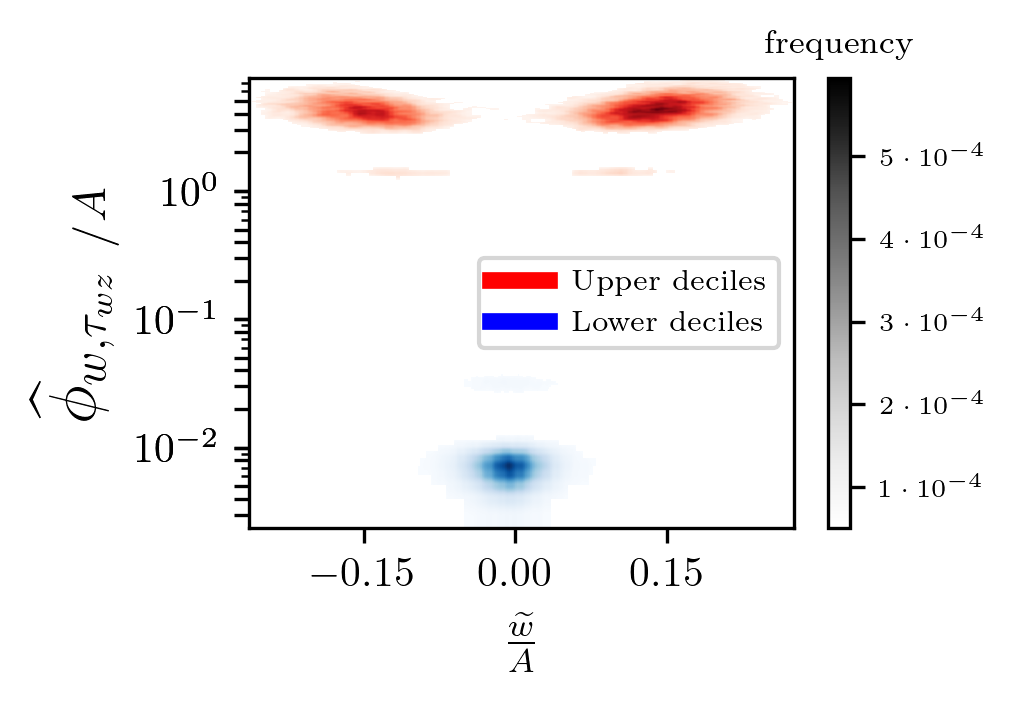}
\caption{Probability of the importance clusters for predicting the streamwise, wall-normal and spanwise velocity from the spanwise shear stress. Joint probability of the averaged physical properties of the structures, averaged spanwise shear stress (left) and average velocity fluctuation (right), and the importance of the pressure for the prediction of streamwise (top), wall-normal (middle) and spanwise (bottom) velocity.}
\label{fig:phyprop_1}
\end{figure}

The physical properties of the structures resulting from predicting each velocity fluctuation from each wall measurement are presented in Figures \ref{fig:phyprop_2}, \ref{fig:phyprop_0} and \ref{fig:phyprop_1}. Figure \ref{fig:phyprop_2} shows the physical properties of the low- and high-importance structures for predicting wall-normal and spanwise velocity from wall-pressure measurements. The high-importance structures exhibit a simetrical distribution of the averaged wall pressure, similar to the streamwise fluctuation of the velocity, Figure \ref{fig:bincount_15}. Two high-importance regions are detected, and as in the case of the streamwise velocity, the high-importance regions are obtained at one to two standard deviations from the averaged value of the pressure for the prediction of the wall-normal and the spanwise velocity fluctuations. In addition, the wall-normal velocity of the high-importance regions is centered, showing a major similarity with the streamwise streaks~\citep{kli67} than with the Reynolds-stress structures~\citep{Lozano2012}. This idea is corroborated by the spanwise velocity, being the high-importance structures located for positive and negative values of the spanwise fluctuations. Furthermore, the maximum presence of streaks is located at a wall-normal distance of $y^+\approx 15$, reinforcing the idea of high-importance streaks.

Figure \ref{fig:phyprop_0} shows the physical properties of the streamwise shear-stress structures for the prediction of the streamwise, wall-normal and spanwise velocity. The figures evidence that the high-influence structures are mostly related to low and high values of the shear stress. The distribution is skewed for the three velocity fluctuation components. For the streamwise velocity prediction, most of the structures are obtained for low values of the streamwise shear stress, while in the case of the wall-normal velocity, the highest probability is obtained for high values of the shear stress. The distribution of the spanwise fluctuation obtains high probabilities for both low and high values of the shear stress. The clusters of importance of the streamwise shear stress are related to positive and negative streamwise and spanwise velocity fluctuations, reinforcing the idea of high-importance streaks, but they exhibit a clear correlation with negative wall-normal fluctuations. This idea links the shear stress with the intense Reynolds-stress events~\citep{Lozano2012} and points to the sweep-like structures, regions with positive streamwise fluctuations and negative wall-normal fluctuations, as the most relevant regions for the wall friction (streamwise shear stress). These results are in agreement with \citet{osawa2024}, who stated that the sweep-like structures are the most influential for the flow evolution.

Figure \ref{fig:phyprop_1} evidences that spanwise wall shear stress high-intensity regions are essential for the prediction of the velocity fluctuations. These regions are mostly related to the positive streamwise velocity and negative wall-normal velocity, reinforcing again the idea of the sweep-like structures influencing the shear stress. In addition, as happened with the streamwise wall-shear stress, the spanwise wall-shear stress clusters are associated with high absolute values of the spanwise velocity, suggesting that part of these regions are associated with the streamwise streaks.

%Finally, A deeper explanation of the properties of the high and low-importance structures for the different pairs input-output is presented in Appendix \ref{appendix:phys}.

\section{Conclusion}\label{sec:conclusions}

The present work has provided an explainable framework of the previous model to interpret and understand the complex relationships between the inputs and the outputs generated in a deep-learning-based flow estimation.

Firstly, the relationships between the input field values and their respective importance scores were presented. The results evidenced that the pressure was the most influential input, as it is the one with the largest SHAP values for all the velocity fluctuations. This idea means that the present estimator is more sensitive to the pressure, and thus, missing or corrupted wall-pressure measurements become critical for the prediction of the velocity fluctuations. Then, streamwise and spanwise wall-shear stresses exhibit similar scores for the streamwise velocity, while in the case of the spanwise and wall-normal velocity, the influence of the streamwise wall-shear stress is lower.

%The wall-normal velocity is evidenced to be the less influenced component as its associated SHAP values present lower absolute scores. This fact is consistent with the idea of streaks, as at a normal-wall distance of $15$ wall units the presence of streaks is maximum and the streaks depend on the combination of the streamwise and spanwise fluctuations.

In addition, the correlation between the high-importance regions and the regions of highest wall-shear-stress and wall-pressure magnitude has been shown to be low. The analysis also included the correlation between the high-importance and high-velocity fluctuation regions and between the high wall-shear-stress and the high-velocity fluctuations regions, showing similar trends, evidencing the ability of the additive-feature-attribution methods to identify the relationships generated by the model instead of quantifying the intensity of the input variables. Indeed, the analysis showed that the regions with the top 20\% importance are the ones with the highest impact on the error of the output and the rest do not significantly affect the predictions. The effect of the wall pressure on the predictions is evidenced as the field becomes non-informative and the error decreases when approximately half of the input grid points are removed.

Finally, the grid points were clustered by dividing the domain into structures of similar importance. Then, the clusters were filtered using a minimum area of $\left(30\right)^2$ viscous units, evidencing that only the top and bottom 10\% of the points based on their importance can form structures large enough to pass the filter. These clusters are strongly related to their average pressure along their area, and they exhibit high importance for an averaged pressure one to two standard deviations from the mean.

Therefore, the present analysis has provided insights into the important regions for predicting the flow through non-intrusive sensing. The results evidenced that the wall pressure is the most influential magnitude. The higher SHAP values associated with the pressure indicate a higher sensitivity of the model to incorrect or inaccurate pressure data. Therefore, the accuracy of the pressure sensors should be prioritized to increase the accuracy of the flow reconstruction. Then, only those regions with a pressure between one and two standard deviations from the mean are significant. These high-importance wall regions for the reconstruction of the flow exhibit a typical size of approximately $80\times40$ wall units, including, inside the importance clusters, 6 grid points in each direction as average for the present analysis. The expected size of the high-importance regions is a requirement in order to obtain enough resolution to define the important clusters in an experimental case. The size of the clusters was compared with the typical equivalence for a commercial flight, showing that with the standard size of wall-stress sensors, the resolution will not be enough to satisfy the criteria required for measuring all the structures of the flow, and further research must be conducted exploring the reproduction of high resolution flows from coarser grids.

\backsection[Aknowledgements and funding]{R.V. acknowledges financial support from ERC grant no. ‘2021-CoG-101043998, DEEPCONTROL’. Views and opinions expressed are however those of the author(s) only and do not necessarily reflect those of the European Union or the European Research Council. Neither the European Union nor the granting authority can be held responsible for them. Part of the deep-learning-model training was enabled by resources provided by the National Academic Infrastructure for Supercomputing in Sweden (NAISS) at Berzelius (NSC), partially funded by the Swedish Research Council through grant agreement no. 2022-06725. S.H. has the support of grant PID2021-128676OB-I00 funded by MCIN/AEI/10.13039/501100011033 and by “ERDF A way of making Europe”, by the European Union. A.I. and S.D. were supported by the project EXCALIBUR (Grant No PID2022-138314NB-I00), funded by MCIU/AEI/ 10.13039/501100011033 and by“ERDF A way of making Europe”.}

\backsection[Data availability statement]{The data and code that support the findings of this study are openly available at http://doi.org/[doi]}

\backsection[Author ORCIDs]{A. Cremades, https://orcid.org/0000-0002-7052-4913; R. Freibergs, https://orcid.org/0009-0000-6360-5603; S. Hoyas, https://orcid.org/0000-0002-8458-7288; A. Ianiro, https://orcid.org/0000-0001-7342-4814; S. Discetti, https://orcid.org/0000-0001-9025-1505; R. Vinuesa,
https://orcid.org/0000-0001-6570-5499}

\backsection[Author contributions]{A. Cremades: Methodology, Writing - Original Draft, Visualization, Validation, Formal Analysis, Investigation, Data Curation. R. Freibergs: Software, Investigation, Visualization, Validation, Formal Analysis, Data Curation. S. Hoyas: Writing - Review \& Editing. A. Ianiro: Writing - Review \& Editing, Data Curation, Conceptualization, Methodology S. Discetti: Writing - Review \& Editing, Data Curation, Conceptualization, Methodology. R. Vinuesa: Supervision, Writing - Review \& Editing, Project administration, Funding acquisition, Conceptualization, Methodology, Resources}

\backsection[Declaration of Interests]{The authors report no conflict of interest.}

\bibliographystyle{jfm}
\bibliography{jfm}

\providecommand{\noopsort}[1]{}\providecommand{\singleletter}[1]{#1}
\begin{thebibliography}{58}
\expandafter\ifx\csname natexlab\endcsname\relax\def\natexlab#1{#1}\fi
\def\au#1{#1} \def\ed#1{#1} \def\yr#1{#1}\def\at#1{#1}\def\jt#1{\textit{#1}} \def\bt#1{#1}\def\bvol#1{\textbf{#1}} \def\vol#1{#1} \def\pg#1{#1} \def\publ#1{#1}\def\arxiv#1{#1}\def\org#1{#1}\def\st#1{\textit{#1}}

\bibitem[Adrian \& Westerweel(2011)]{adrian2011}
{\sc \au{Adrian, Ronald~J} \& \au{Westerweel, Jerry}} \yr{2011} {\em Particle image velocimetry\/}.  \publ{Cambridge university press}.

\bibitem[Agency(2020)]{IEA2020}
{\sc \au{Agency, International~Energy}} \yr{2020}  \bt{Key world energy statistics, https://www.iea.org/reports/key-world-energy-statistics-2020, accessed 20-nov-2022}. IEA, Paris.

\bibitem[Al-Jarrah {\em et~al.\/}(2015)Al-Jarrah, Yoo, Muhaidat, Karagiannidis \& Taha]{al2015}
{\sc \au{Al-Jarrah, Omar~Y}, \au{Yoo, Paul~D}, \au{Muhaidat, Sami}, \au{Karagiannidis, George~K} \& \au{Taha, Kamal}} \yr{2015}  \at{Efficient machine learning for big data: A review}.  \jt{Big Data Research}  \bvol{2}~(3),  \pg{87--93}.

\bibitem[Bar-Sinai {\em et~al.\/}(2019)Bar-Sinai, Hoyer, Hickey \& Brenner]{bar2019}
{\sc \au{Bar-Sinai, Yohai}, \au{Hoyer, Stephan}, \au{Hickey, Jason} \& \au{Brenner, Michael~P}} \yr{2019}  \at{Learning data-driven discretizations for partial differential equations}.  \jt{Proceedings of the National Academy of Sciences}  \bvol{116}~(31),  \pg{15344--15349}.

\bibitem[Beck {\em et~al.\/}(2019)Beck, Flad \& Munz]{beck2019}
{\sc \au{Beck, Andrea}, \au{Flad, David} \& \au{Munz, Claus-Dieter}} \yr{2019}  \at{Deep neural networks for data-driven les closure models}.  \jt{Journal of Computational Physics}  \bvol{398},  \pg{108910}.

\bibitem[Boussinesq(1903)]{boussinesq1903}
{\sc \au{Boussinesq, Joseph}} \yr{1903} {\em Th{\'e}orie analytique de la chaleur: mise en harmonie avec la thermodynamique et avec la th{\'e}orie m{\'e}canique de la lumi{\`e}re\/}, ,  \vol{vol.~2}.  \publ{Gauthier-Villars}.

\bibitem[Cardesa {\em et~al.\/}(2017)Cardesa, Vela-Mart\'in \& Jim\'enez]{Cardesa_science}
{\sc \au{Cardesa, José~I.}, \au{Vela-Mart\'in, Alberto} \& \au{Jim\'enez, Javier}} \yr{2017}  \at{The turbulent cascade in five dimensions}.  \jt{Science}  \bvol{357},  \pg{782--784}.

\bibitem[Chui {\em et~al.\/}(2018)Chui, Manyika, Miremadi, Henke, Chung, Nel \& Malhotra]{chui2018}
{\sc \au{Chui, Michael}, \au{Manyika, James}, \au{Miremadi, Mehdi}, \au{Henke, Nicolaus}, \au{Chung, Rita}, \au{Nel, Pieter} \& \au{Malhotra, Sankalp}} \yr{2018}  \at{Notes from the ai frontier: Insights from hundreds of use cases}.  \jt{McKinsey Global Institute}  \bvol{2},  \pg{267}.

\bibitem[Cremades {\em et~al.\/}(2024)Cremades, Hoyas, Deshpande, Quintero, Lellep, Lee, Monty, Hutchins, Linkmann, Marusic {\em et~al.\/}]{cremades2024}
{\sc \au{Cremades, Andr{\'e}s}, \au{Hoyas, Sergio}, \au{Deshpande, Rahul}, \au{Quintero, Pedro}, \au{Lellep, Martin}, \au{Lee, Will~Junghoon}, \au{Monty, Jason~P}, \au{Hutchins, Nicholas}, \au{Linkmann, Moritz}, \au{Marusic, Ivan} \& \au{others}} \yr{2024}  \at{Identifying regions of importance in wall-bounded turbulence through explainable deep learning}.  \jt{Nature Communications}  \bvol{15}~(1),  \pg{3864}.

\bibitem[Cremades {\em et~al.\/}(2025)Cremades, Hoyas \& Vinuesa]{cremades2025}
{\sc \au{Cremades, Andr{\'e}s}, \au{Hoyas, Sergio} \& \au{Vinuesa, Ricardo}} \yr{2025}  \at{Additive-feature-attribution methods: a review on explainable artificial intelligence for fluid dynamics and heat transfer}.  \jt{International Journal of Heat and Fluid Flow}  \bvol{112},  \pg{109662}.

\bibitem[Cu{\'e}llar {\em et~al.\/}(2024{\natexlab{{\em a\/}}})Cu{\'e}llar, G{\"u}emes, Ianiro, Flores, Vinuesa \& Discetti]{cuellar2024}
{\sc \au{Cu{\'e}llar, Antonio}, \au{G{\"u}emes, Alejandro}, \au{Ianiro, Andrea}, \au{Flores, {\'O}scar}, \au{Vinuesa, Ricardo} \& \au{Discetti, Stefano}} \yr{2024{\natexlab{{\em a\/}}}}  \at{Three-dimensional generative adversarial networks for turbulent flow estimation from wall measurements}.  \jt{Journal of Fluid Mechanics}  \bvol{991},  \pg{A1}.

\bibitem[Cu{\'e}llar {\em et~al.\/}(2024{\natexlab{{\em b\/}}})Cu{\'e}llar, Ianiro \& Discetti]{cuellar2024_2}
{\sc \au{Cu{\'e}llar, Antonio}, \au{Ianiro, Andrea} \& \au{Discetti, Stefano}} \yr{2024{\natexlab{{\em b\/}}}}  \at{Some effects of limited wall-sensor availability on flow estimation with 3d-gans}.  \jt{Theoretical and Computational Fluid Dynamics}  \bvol{38}~(5),  \pg{729--745}.

\bibitem[Deshpande {\em et~al.\/}(2023)Deshpande, Van Den~Bogaard, Vinuesa, Lindi{\'c} \& Marusic]{deshpande2023}
{\sc \au{Deshpande, Rahul}, \au{Van Den~Bogaard, Aron}, \au{Vinuesa, Ricardo}, \au{Lindi{\'c}, Luka} \& \au{Marusic, Ivan}} \yr{2023}  \at{Reynolds-number effects on the outer region of adverse-pressure-gradient turbulent boundary layers}.  \jt{Physical Review Fluids}  \bvol{8}~(12),  \pg{124604}.

\bibitem[Ganapathisubramani {\em et~al.\/}(2005)Ganapathisubramani, Longmire, Marusic \& Pothos]{ganapathisubramani2005}
{\sc \au{Ganapathisubramani, Bharathram}, \au{Longmire, Ellen~K}, \au{Marusic, Ivan} \& \au{Pothos, Stamatios}} \yr{2005}  \at{Dual-plane piv technique to determine the complete velocity gradient tensor in a turbulent boundary layer}.  \jt{Experiments in Fluids}  \bvol{39},  \pg{222--231}.

\bibitem[George \& Lumley(1973)]{george1973}
{\sc \au{George, William~K} \& \au{Lumley, John~L}} \yr{1973}  \at{The laser-doppler velocimeter and its application to the measurement of turbulence}.  \jt{Journal of Fluid Mechanics}  \bvol{60}~(2),  \pg{321--362}.

\bibitem[Guastoni {\em et~al.\/}(2021)Guastoni, Güemes, Ianiro, Discetti, Schlatter, Azizpour \& Vinuesa]{Guastoni2020}
{\sc \au{Guastoni, Luca}, \au{Güemes, Alejandro}, \au{Ianiro, Andrea}, \au{Discetti, Stefano}, \au{Schlatter, Philipp}, \au{Azizpour, Hossein} \& \au{Vinuesa, Ricardo}} \yr{2021}  \at{Convolutional-network models to predict wall-bounded turbulence from wall quantities}.  \jt{Journal of Fluid Mechanics}  \bvol{928},  \pg{A27}.

\bibitem[H{\aa}kansson {\em et~al.\/}(2013)H{\aa}kansson, Kvick, Lundell, Prahl~Wittberg \& S{\"o}derberg]{haakansson2013}
{\sc \au{H{\aa}kansson, Karl~MO}, \au{Kvick, Mathias}, \au{Lundell, Fredrik}, \au{Prahl~Wittberg, Lisa} \& \au{S{\"o}derberg, L~Daniel}} \yr{2013}  \at{Measurement of width and intensity of particle streaks in turbulent flows}.  \jt{Experiments in fluids}  \bvol{54},  \pg{1--13}.

\bibitem[He {\em et~al.\/}(2022)He, Tan, Rigas \& Vahdati]{he2022}
{\sc \au{He, Xiao}, \au{Tan, Jianheng}, \au{Rigas, Georgios} \& \au{Vahdati, Mehdi}} \yr{2022}  \at{On the explainability of machine-learning-assisted turbulence modeling for transonic flows}.  \jt{International Journal of Heat and Fluid Flow}  \bvol{97},  \pg{109038}.

\bibitem[Hoyas {\em et~al.\/}(2022)Hoyas, Oberlack, Alc\'antara-\'Avila, Kraheberger \& Laux]{hoy22}
{\sc \au{Hoyas, Sergio}, \au{Oberlack, Martin}, \au{Alc\'antara-\'Avila, Francisco}, \au{Kraheberger, Stefanie~V.} \& \au{Laux, Jonathan}} \yr{2022}  \at{{Wall turbulence at high friction Reynolds numbers}}.  \jt{Physical Review Fluids}  \bvol{7},  \pg{014602}.

\bibitem[I{\~n}igo {\em et~al.\/}(2014)I{\~n}igo, Sipp \& Schmid]{inigo2014}
{\sc \au{I{\~n}igo, Juan~Guzm{\'a}n}, \au{Sipp, Denis} \& \au{Schmid, Peter~J}} \yr{2014}  \at{A dynamic observer to capture and control perturbation energy in noise amplifiers}.  \jt{Journal of fluid mechanics}  \bvol{758},  \pg{728--753}.

\bibitem[Jia {\em et~al.\/}(2019)Jia, Dao, Wang, Hubis, Hynes, G{\"u}rel, Li, Zhang, Song \& Spanos]{jia2019}
{\sc \au{Jia, Ruoxi}, \au{Dao, David}, \au{Wang, Boxin}, \au{Hubis, Frances~Ann}, \au{Hynes, Nick}, \au{G{\"u}rel, Nezihe~Merve}, \au{Li, Bo}, \au{Zhang, Ce}, \au{Song, Dawn} \& \au{Spanos, Costas~J}} \yr{2019} Towards efficient data valuation based on the shapley value.  \bt{In {\em The 22nd International Conference on Artificial Intelligence and Statistics\/}},  \pg{pp. 1167--1176}. PMLR.

\bibitem[Jiménez(2013)]{Jimenez2013}
{\sc \au{Jiménez, Javier}} \yr{2013}  \at{Near-wall turbulence}.  \jt{Physics of Fluids}  \bvol{25},  \pg{101302}.

\bibitem[Jiménez(2018)]{Jimenez2018}
{\sc \au{Jiménez, Javier}} \yr{2018}  \at{Coherent structures in wall-bounded turbulence}.  \jt{Journal of Fluid Mechanics}  \bvol{842},  \pg{P1}.

\bibitem[Kaiser {\em et~al.\/}(2014)Kaiser, Noack, Cordier, Spohn, Segond, Abel, Daviller, {\"O}sth, Krajnovi{\'c} \& Niven]{kaiser2014}
{\sc \au{Kaiser, Eurika}, \au{Noack, Bernd~R}, \au{Cordier, Laurent}, \au{Spohn, Andreas}, \au{Segond, Marc}, \au{Abel, Markus}, \au{Daviller, Guillaume}, \au{{\"O}sth, Jan}, \au{Krajnovi{\'c}, Sini{\v{s}}a} \& \au{Niven, Robert~K}} \yr{2014}  \at{Cluster-based reduced-order modelling of a mixing layer}.  \jt{Journal of Fluid Mechanics}  \bvol{754},  \pg{365--414}.

\bibitem[Kim(2011)]{kim2011}
{\sc \au{Kim, John}} \yr{2011}  \at{Physics and control of wall turbulence for drag reduction}.  \jt{Philosophical Transactions of the Royal Society A: Mathematical, Physical and Engineering Sciences}  \bvol{369}~(1940),  \pg{1396--1411}.

\bibitem[Kim \& Lee(2020)]{kim2020}
{\sc \au{Kim, Junhyuk} \& \au{Lee, Changhoon}} \yr{2020}  \at{Prediction of turbulent heat transfer using convolutional neural networks}.  \jt{Journal of Fluid Mechanics}  \bvol{882},  \pg{A18}.

\bibitem[Kim {\em et~al.\/}(2024)Kim, Kim, Song \& You]{kim2024}
{\sc \au{Kim, Taewan}, \au{Kim, Changwook}, \au{Song, Jeonghwan} \& \au{You, Donghyun}} \yr{2024}  \at{Optimal control of a wind farm in time-varying wind using deep reinforcement learning}.  \jt{Energy}  \pg{p. 131950}.

\bibitem[Kline {\em et~al.\/}(1967)Kline, Reynolds, Schraub \& Runstadler]{kli67}
{\sc \au{Kline, S.~J.}, \au{Reynolds, W.~C.}, \au{Schraub, F.~A.} \& \au{Runstadler, P.~W.}} \yr{1967}  \at{The structure of turbulent boundary layers}.  \jt{Journal of Fluid Mechanics}  \bvol{30}~(4),  \pg{741–773}.

\bibitem[Kochkov {\em et~al.\/}(2021)Kochkov, Smith, Alieva, Wang, Brenner \& Hoyer]{kochkov2021}
{\sc \au{Kochkov, Dmitrii}, \au{Smith, Jamie~A}, \au{Alieva, Ayya}, \au{Wang, Qing}, \au{Brenner, Michael~P} \& \au{Hoyer, Stephan}} \yr{2021}  \at{Machine learning--accelerated computational fluid dynamics}.  \jt{Proceedings of the National Academy of Sciences}  \bvol{118}~(21),  \pg{e2101784118}.

\bibitem[Kolmogorov(1941)]{kol41a}
{\sc \au{Kolmogorov, A.~N.}} \yr{1941}  \at{{Local structure of turbulence in an incompressible fluid at very high Reynolds numbers.}}  \jt{Dokl. Akad. Nauk.}  \bvol{SSSR (30)},  \pg{9--13}.

\bibitem[LeCun {\em et~al.\/}(2015)LeCun, Bengio \& Hinton]{lecun2015}
{\sc \au{LeCun, Yann}, \au{Bengio, Yoshua} \& \au{Hinton, Geoffrey}} \yr{2015}  \at{Deep learning}.  \jt{Nature}  \bvol{521}~(7553),  \pg{436--444}.

\bibitem[LeCun {\em et~al.\/}(1989)LeCun, Boser, Denker, Henderson, Howard, Hubbard \& Jackel]{lecun1989}
{\sc \au{LeCun, Yann}, \au{Boser, Bernhard}, \au{Denker, John~S}, \au{Henderson, Donnie}, \au{Howard, Richard~E}, \au{Hubbard, Wayne} \& \au{Jackel, Lawrence~D}} \yr{1989}  \at{Backpropagation applied to handwritten zip code recognition}.  \jt{Neural computation}  \bvol{1}~(4),  \pg{541--551}.

\bibitem[Lee {\em et~al.\/}(1997)Lee, Kim, Babcock \& Goodman]{lee1997}
{\sc \au{Lee, Changhoon}, \au{Kim, John}, \au{Babcock, David} \& \au{Goodman, Rodney}} \yr{1997}  \at{Application of neural networks to turbulence control for drag reduction}.  \jt{Physics of Fluids}  \bvol{9}~(6),  \pg{1740--1747}.

\bibitem[Lee \& Carlberg(2020)]{lee2020}
{\sc \au{Lee, Kookjin} \& \au{Carlberg, Kevin~T}} \yr{2020}  \at{Model reduction of dynamical systems on nonlinear manifolds using deep convolutional autoencoders}.  \jt{Journal of Computational Physics}  \bvol{404},  \pg{108973}.

\bibitem[Lellep {\em et~al.\/}(2022)Lellep, Prexl, Eckhardt \& Linkmann]{Lellep2022}
{\sc \au{Lellep, Martin}, \au{Prexl, Jonathan}, \au{Eckhardt, Bruno} \& \au{Linkmann, Moritz}} \yr{2022}  \at{Interpreted machine learning in fluid dynamics: explaining relaminarisation events in wall-bounded shear flows}.  \jt{Journal of Fluid Mechanics}  \bvol{942},  \pg{A2}.

\bibitem[Li {\em et~al.\/}(2020)Li, Kovachki, Azizzadenesheli, Liu, Bhattacharya, Stuart \& Anandkumar]{li2020}
{\sc \au{Li, Zongyi}, \au{Kovachki, Nikola}, \au{Azizzadenesheli, Kamyar}, \au{Liu, Burigede}, \au{Bhattacharya, Kaushik}, \au{Stuart, Andrew} \& \au{Anandkumar, Anima}} \yr{2020}  \at{Fourier neural operator for parametric partial differential equations, arxiv}.  \jt{preprint arXiv:2010.08895} .

\bibitem[Liu {\em et~al.\/}(1991)Liu, Landreth, Adrian \& Hanratty]{liu1991}
{\sc \au{Liu, Z-C}, \au{Landreth, CC}, \au{Adrian, RJ} \& \au{Hanratty, TJ}} \yr{1991}  \at{High resolution measurement of turbulent structure in a channel with particle image velocimetry}.  \jt{Experiments in fluids}  \bvol{10},  \pg{301--312}.

\bibitem[Lozano-Durán {\em et~al.\/}(2012)Lozano-Durán, Flores \& Jiménez]{Lozano2012}
{\sc \au{Lozano-Durán, Adrián}, \au{Flores, Oscar} \& \au{Jiménez, Javier}} \yr{2012}  \at{The three-dimensional structure of momentum transfer in turbulent channels}.  \jt{Journal of Fluid Mechanics}  \bvol{694},  \pg{100--130}.

\bibitem[Lundberg \& Lee(2017)]{lundberg2017}
{\sc \au{Lundberg, S.~M.} \& \au{Lee, S.}} \yr{2017}  \at{A unified approach to interpreting model predictions}.  \jt{Advances in neural information processing systems}  \bvol{30}.

\bibitem[Marino {\em et~al.\/}(2024)Marino, Juanicotena, Errasti, Mayoral, Manrique~de Lara, Vinuesa \& Ferrer]{marino2024}
{\sc \au{Marino, Oscar~A}, \au{Juanicotena, Adrian}, \au{Errasti, Jon}, \au{Mayoral, David}, \au{Manrique~de Lara, Fernando}, \au{Vinuesa, Ricardo} \& \au{Ferrer, Esteban}} \yr{2024}  \at{A comparison of neural-network architectures to accelerate high-order h/p solvers}.  \jt{Physics of Fluids}  \bvol{36}~(10).

\bibitem[Marusic {\em et~al.\/}(2021)Marusic, Chandran, Rouhi, Fu, Wine, Holloway, Chung \& Smits]{marusic2021}
{\sc \au{Marusic, Ivan}, \au{Chandran, Dileep}, \au{Rouhi, Amirreza}, \au{Fu, Matt~K}, \au{Wine, David}, \au{Holloway, Brian}, \au{Chung, Daniel} \& \au{Smits, Alexander~J}} \yr{2021}  \at{An energy-efficient pathway to turbulent drag reduction}.  \jt{Nature communications}  \bvol{12}~(1),  \pg{5805}.

\bibitem[McConkey {\em et~al.\/}(2022)McConkey, Yee \& Lien]{mcconkey2022}
{\sc \au{McConkey, Ryley}, \au{Yee, Eugene} \& \au{Lien, Fue-Sang}} \yr{2022}  \at{Deep structured neural networks for turbulence closure modeling}.  \jt{Physics of Fluids}  \bvol{34}~(3).

\bibitem[Morimoto {\em et~al.\/}(2021)Morimoto, Fukami, Zhang, Nair \& Fukagata]{morimoto2021}
{\sc \au{Morimoto, Masaki}, \au{Fukami, Kai}, \au{Zhang, Kai}, \au{Nair, Aditya~G} \& \au{Fukagata, Koji}} \yr{2021}  \at{Convolutional neural networks for fluid flow analysis: toward effective metamodeling and low dimensionalization}.  \jt{Theoretical and Computational Fluid Dynamics}  \bvol{35}~(5),  \pg{633--658}.

\bibitem[Murata {\em et~al.\/}(2020)Murata, Fukami \& Fukagata]{murata2020}
{\sc \au{Murata, Takaaki}, \au{Fukami, Kai} \& \au{Fukagata, Koji}} \yr{2020}  \at{Nonlinear mode decomposition with convolutional neural networks for fluid dynamics}.  \jt{Journal of Fluid Mechanics}  \bvol{882},  \pg{A13}.

\bibitem[Navier(1827)]{navier1827}
{\sc \au{Navier, CL}} \yr{1827}  \at{Memoir on the laws of fluid motion}.  \jt{Mem., acad. sci.(Paris)}  \bvol{6},  \pg{389}.

\bibitem[Osawa \& Jiménez(2024)]{osawa2024}
{\sc \au{Osawa, Kosuke} \& \au{Jiménez, Javier}} \yr{2024}  \at{Causal features in turbulent channel flow}.  \jt{Journal of Fluid Mechanics}  \bvol{1000},  \pg{A4}.

\bibitem[Rabault {\em et~al.\/}(2017)Rabault, Kolaas \& Jensen]{rabault2017}
{\sc \au{Rabault, Jean}, \au{Kolaas, Jostein} \& \au{Jensen, Atle}} \yr{2017}  \at{Performing particle image velocimetry using artificial neural networks: a proof-of-concept}.  \jt{Measurement Science and Technology}  \bvol{28}~(12),  \pg{125301}.

\bibitem[Ribeiro {\em et~al.\/}(2016)Ribeiro, Singh \& Guestrin]{Ribeiro2016}
{\sc \au{Ribeiro, Marco~Tulio}, \au{Singh, Sameer} \& \au{Guestrin, Carlos}} \yr{2016}  \bt{{Why Should I Trust You? Explaining the Predictions of Any Classifier}}.  \pg{pp. 1135--1144}. In Proceedings of the 22nd ACM SIGKDD International Conference on Knowledge Discovery and Data Mining.

\bibitem[Shapley(1953)]{shapley1953}
{\sc \au{Shapley, Lloyd~S}} \yr{1953}  \at{{A value for n-person games}}.  \jt{Contribution to the Theory of Games}  \bvol{2}.

\bibitem[Stokes(2009)]{Stokes2009}
{\sc \au{Stokes, George~Gabriel}} \yr{2009} {\em On the Theories of the Internal Friction of Fluids in Motion, and of the Equilibrium and Motion of Elastic Solids\/},  \pg{p. 75–129}. {\em Cambridge Library Collection - Mathematics\/} .  \publ{Cambridge University Press}.

\bibitem[Sudharsun \& Warrior(2023)]{sudharsun2023}
{\sc \au{Sudharsun, G} \& \au{Warrior, Hari~V}} \yr{2023}  \at{Enhancing turbulence modeling: Machine learning for pressure-strain correlation and uncertainty quantification in the reynolds stress model}.  \jt{Physics of Fluids}  \bvol{35}~(12).

\bibitem[Taylor(1938)]{tay38}
{\sc \au{Taylor, G.~I.}} \yr{1938}  \at{The spectrum of turbulence}.  \jt{Proceedings of the Royal Society of London. Series A - Mathematical and Physical Sciences}  \bvol{164}~(919),  \pg{476--490}.

\bibitem[Townsend(1976)]{tow76}
{\sc \au{Townsend, A.A.}} \yr{1976} {\em The Structure of Turbulent Shear Flows, 2nd ed.\/}.  \publ{New York: Cambridge University Press}.

\bibitem[Vinuesa \& Brunton(2022)]{brunton}
{\sc \au{Vinuesa, R.} \& \au{Brunton, S.L.}} \yr{2022}  \at{Enhancing computational fluid dynamics with machine learning}.  \jt{Nature Computational Science}  \bvol{2},  \pg{358--366}.

\bibitem[Vinuesa {\em et~al.\/}(2023)Vinuesa, Brunton \& McKeon]{vinuesa2023}
{\sc \au{Vinuesa, Ricardo}, \au{Brunton, Steven~L} \& \au{McKeon, Beverley~J}} \yr{2023}  \at{The transformative potential of machine learning for experiments in fluid mechanics}.  \jt{Nature Reviews Physics}  \bvol{5}~(9),  \pg{536--545}.

\bibitem[Wang {\em et~al.\/}(2017)Wang, Wu \& Xiao]{wang2017}
{\sc \au{Wang, Jian-Xun}, \au{Wu, Jin-Long} \& \au{Xiao, Heng}} \yr{2017}  \at{Physics-informed machine learning approach for reconstructing reynolds stress modeling discrepancies based on dns data}.  \jt{Physical Review Fluids}  \bvol{2}~(3),  \pg{034603}.

\bibitem[Yousif {\em et~al.\/}(2023)Yousif, Yu, Hoyas, Vinuesa \& Lim]{yousif2023}
{\sc \au{Yousif, Mustafa~Z}, \au{Yu, Linqi}, \au{Hoyas, Sergio}, \au{Vinuesa, Ricardo} \& \au{Lim, HeeChang}} \yr{2023}  \at{A deep-learning approach for reconstructing 3d turbulent flows from 2d observation data}.  \jt{Scientific Reports}  \bvol{13}~(1),  \pg{2529}.

\bibitem[Zhe {\em et~al.\/}(2005)Zhe, Modi \& Farmer]{zhe2005}
{\sc \au{Zhe, Jiang}, \au{Modi, Vijay} \& \au{Farmer, Kenneth~R}} \yr{2005}  \at{A microfabricated wall shear-stress sensor with capacitative sensing}.  \jt{Journal of Microelectromechanical Systems}  \bvol{14}~(1),  \pg{167--175}.

\end{thebibliography}
%Use of the above commands will create a bibliography using the .bib file. Shown below is a bibliography built from individual items.

%\bibliographystyle{jfm}
%\bibliography{jfm2esam}

%% End of file `jfm2esam.bib'.

\end{document}